\documentclass[a4paper,11pt]{article}

\usepackage{xypic}

\usepackage{jheppub,graphicx,floatflt}
\usepackage{amsmath,latexsym,amssymb,slashed}
\usepackage{enumerate}
\usepackage{mathdots}
\usepackage{mathtools}
\usepackage{extarrows}
\usepackage{mathrsfs}
\usepackage[colorlinks=true,%
            linkcolor=blue,%
            citecolor=blue,%
            urlcolor=blue]{hyperref}

\usepackage{dsfont}
\usepackage{multirow}
\newcommand\beal{\begin{align}}
\newcommand\nn{\nonumber}

\newcommand{\eq}[1]{\begin{equation}#1\end{equation}}
\newcommand{\spl}[1]{\begin{split}#1\end{split}}

\newcommand{\mcal}{\mathcal{M}}

\newcommand{\g}{\gamma}

\renewcommand{\O}{\Omega}

\newcommand{\boxedeq}[1]{
\begin{equation}
\fbox{
\rule[0.7cm]{0pt}{0pt}
$#1$
\rule[-0.45cm]{0pt}{0pt}
}
\end{equation}
}

\def\d{\text{d}}

\def\slashchar#1{\setbox0=\hbox{$#1$}           
\dimen0=\wd0                                 
\setbox1=\hbox{/} \dimen1=\wd1               
\ifdim\dimen0>\dimen1                        
\rlap{\hbox to \dimen0{\hfil/\hfil}}      
#1                                        
\else                                        
\rlap{\hbox to \dimen1{\hfil$#1$\hfil}}   
/                                         
\fi}

\usepackage{xcolor}
\usepackage{color}

\xdefinecolor{mygreen}{rgb}{.2,.55,.2} 

\xdefinecolor{myred}{rgb}{.5,.2,.2} 

\xdefinecolor{mygrey}{rgb}{.8,.8,.8}  

\xdefinecolor{myblue}{rgb}{.15,.15,.45}  

\xdefinecolor{myblack}{rgb}{.27,.27,.27}

\newcommand{\cnote}[1]{}

\title{Generalized geometry lectures on type II backgrounds}

\author{Dimitrios Tsimpis}

\affiliation{
Universit\'e Claude Bernard (Lyon 1)\\
CNRS/IN2P3, UMR5822, Institut de Physique Nucl\'eaire de Lyon\\
   69622 Villeurbanne Cedex, France}

\emailAdd{tsimpis@ipnl.in2p3.fr}

\abstract{
The first part of these notes is a self-contained introduction to generalized complex geometry. 
It is intended as a `user manual' for tools used in the study of supersymmetric backgrounds of supergravity. In the second part 
we review some past and recent results on the generalized complex structure 
of supersymmetric type II vacua in various dimensions. 
} 

\begin{document}
\maketitle
\flushbottom
\setcounter{footnote}{0}
\renewcommand{\thefootnote}{\arabic{footnote}}
\setcounter{section}{0}

\section{Introduction}

These notes on Generalized Complex Geometry (GCG) are based on lectures given at the Corfu Summer School and at the Simons Center for Geometry and Physics. They  are not intended as an exhaustive survey of the subject, but rather as a `user manual' for several technical tools used in 
exploring supersymmetric solutions (vacua) of supergravity. The main target audience is students who are looking for a hands-on, pedagogical exposition of these techniques.

GCG was originally introduced by Hitchin in \cite{hitch} and further developed by Gualtieri in \cite{gual}. There are by now several excellent reviews of GCG for physicists \cite{koer}, for  mathematicians \cite{Hitchin:2010qz,Gualtieri:2007ng,caval}, or both \cite{Zabzine:2006uz}. 
The reader should consult these references for a more complete survey of the literature. Here we will adopt a somewhat different line of development, motivating the different structures from the point of view of supersymmetric vacua of type II supergravity.

I will assume familiarity with the properties of spinors and gamma matrices in $D$ dimensions, see e.g.~\cite{Strathdee:1986jr}, but otherwise these notes are self-contained. 
Several passages of the text are highlighted in green: these include proofs and examples worked out in detail, and can be omitted at first reading.

\subsection{Motivation}

Generalized geometry can be viewed as a way to `geometrize' the flux. 
As we will see later on, GCG accomplishes  that for the metric and the $B$-field.\footnote{{\it Exceptional generalized geometry} is a natural extension of this idea to include the remaining fluxes of supegravity. This is however beyond the scope of these notes which focus exclusively on GCG.} 
GCG has proven very useful in searching  for supersymmetric 
solutions of supergravity.  It has also proven very useful in constructing effective actions and consistent truncations. Moreover it has shed new light on old results on supersymmetric sigma models and has stimulated a lot of new activity in this area.

Our motivation here comes from the study of the general structure of supersymmetric type II flux vacua in the light of GCG. More specifically we will examine the possibility of extending the so-called {\it supersymmetry/calibrations correspondence} beyond the case of four external dimensions. 
As we will see there are simplifications which occur for backgrounds admitting certain Killing spinor ans\"{a}tze: these are the so-called {\it pure backgrounds} in which the supersymmetry  parameters of the background are given by (ordinary) pure spinors. As it turns out, pure backgrounds are equivalent to $SU(n)\times SU(n)$-structure backgrounds.

\section{Part I: generalized complex geometry}

In this part we give an introduction to generalized complex geometry. The following is a list of notation for quick reference:

\begin{itemize}
\item $D$ is the internal spacetime dimension; it is assumed even: $D=2n$, $n\in\mathbb{N}$. Since the focus of these notes is on type II supergravity, the maximal value of $D$ is ten. 
\item $n:=D/2$, with $n=0,1,\dots,5$, is half the dimension of the internal spacetime manifold.
\item $d$ is the dimension of external spacetime, $\mathbb{R}^{1,d-1}$, so that $d=10-D$.
\item $\mcal$ is the internal spacetime manifold. It is assumed spin and Riemannian with metric $g$. It has even (real) dimension: $D=2n$. We will also sometimes write $\mcal_{2n}$ to emphasize the dimensionality.
\item Lower-case letters from the middle of the alphabet, $i,j,\dots=1,\dots,D$, are used as indices for the coordinates of $\mcal$. 
\item Upper-case letters from the middle of the alphabet, $M,N,\dots=1,\dots,2D$, are used as indices of sections of the generalized tangent bundle $T\oplus T^*$ of $\mcal$, (i.e.~as indices of generalized vectors).
\item Our spinor conventions are summarized in appendix \ref{sec:spinors}. They are compatible with those of e.g. \cite{Strathdee:1986jr}. 
\end{itemize}

\subsection{Pure spinors}

Consider the even-dimensional, $D=2n$, Euclidean space $\mathbb{R}^{D}$. 
The Clifford algebra $Cl(D)$ is generated by $2n$ gamma matrices: 
\eq{\label{cl}\{\gamma_p,\gamma_q\}=2\delta_{pq}~;~~~p,q=1,\dots,D=2n~.}
It is an associative algebra because (a) it is equipped with an associative product: matrix multiplication, and (b) it is a vector space: 
one may take arbitrary linear combinations of 
products of gamma matrices. 

Alternatively it will be useful to view the Clifford algebra $Cl(D)$ as the associative algebra freely generated by vectors $v\in\mathbb{R}^{D}$ modulo the relations, 
\eq{\label{modulo}v\star w+w\star v=2v\cdot w~,}
where $\star$ is the product of the algebra and $\cdot$ is the ordinary Euclidean scalar product. 
The $2^n\times 2^n$ gamma matrices can then be viewed as providing the least-dimensional faithful representation $\rho$ of $Cl(D)$, so that $\star$ is 
represented by ordinary matrix multiplication.  
Explicitly:
\eq{\label{clal}\rho(v)=v^i\gamma_i~;~~~\rho(v\star w)=v^iw^j\gamma_i\gamma_j~,}
where (\ref{cl}) ensures that the relations (\ref{modulo}) are satisfied. The vector space of the irreducible module $\rho$ is nothing other than the space of (Dirac) spinors, which can be thought of as (generally complex) $2^n$-dimensional column vectors on which the gamma matrices act. We will refer to $\rho(v)$ as the {\it spinorial action of} $v$.

The set of $2n$ gamma matrices can be split into two groups:
\eq{\label{1}\gamma^{\pm}_a:= \displaystyle\frac12(\gamma_{a}\pm i\gamma_{n+a})~;~~~a=1,\dots,n~,}
obeying creation, annihilation commutation relations:
\eq{\label{ca}\{\gamma_a^{\pm},\gamma_b^{\pm}\}=0~;~~~\{\gamma_a^{+},\gamma_b^{-}\}=\delta_{ab}.}
A {\it pure spinor}, $\eta$, is a spinor that is annihilated by exactly 
$n=D/2$ linear combinations of the gamma matrices. Without loss of generality  we may take those to be the 
annihilation operators defined above. 
From this point of view a pure spinor 
is nothing other than the 
Fock vacuum,
\eq{\label{ps}\gamma^{-}_a\eta=0~.}
\color{mygreen}We remark that the pure spinor $\eta$ has positive chirality,
\eq{\label{chirality}\gamma_{D+1}\eta=\eta~,}
in the basis of gamma matrices given by (\ref{cl}),(\ref{ps}), as follows straightforwardly from the definition of the chirality operator,
\eq{\label{2}\gamma_{D+1}=i^n\gamma_1\cdots\gamma_{2n}~.}
Indeed using (\ref{1}) to express the gamma matrices $\gamma_m$, $m=1,\dots,2n$,  in terms of holomorphic, antiholomorphic ones $\gamma_a^{\pm}$, $a=1,\dots,n$, inserting into (\ref{2}) and taking (\ref{ca}),(\ref{ps}) into account, we obtain (\ref{chirality}). Similarly 
the complex conjugate of $\eta$ is defined as,
\eq{\eta^c=C\eta^*~,}
where $C$ on the right hand side above is the charge conjugation matrix: it ensures that $\eta^c$ transforms as a spinor, cf.~appendix \ref{sec:spinors}. 
It follows that $\eta^c$ 
is annihilated by the creation operators,
\eq{\gamma_a^+\eta^c=0~,}
and thus is also a pure spinor (since it is annihilated by $D/2=n$ linear combinations of gamma matrices). Recall that $\eta^c$ has definite chirality: positive (the same as $\eta$) for $n=0,2$ mod 4 and negative for $n=1,3$ mod 4.

\color{black} 

{}Furthermore any Dirac spinor, $\psi$, can be built from the vacuum $\eta$ by acting with linear combinations of products of creation operators: 
\eq{\label{gr}\psi=\sum_{p=0}^n c^{a_1\dots a_p}\gamma_{a_1}^{+}\cdots \gamma_{a_p}^{+}\eta~.}
The coefficients $c^{a_1\dots a_p}$ above are antisymmetric in all $p$ indices, as follows from (\ref{ca}), and so they carry $n$-choose-$p$ degrees of freedom. This implies that the Dirac spinor module has dimension 
\eq{2^n = \sum_{p=0}^n \Big(\begin{array}{c}
n\\
p
\end{array}\Big)~,}
as already mentioned. 
One more point that will be important in the following is that the space of Dirac spinors can be graded by `occupation number', i.e. the number of creation operators that are applied to the Fock vaccum in order to obtain the spinor. Moreover 
spinors with definite  occupation numbers form a basis of the space of Dirac spinors. In other words, an arbitrary Dirac spinor can be thought of as a linear combination of spinors with definite  occupation numbers, cf.~(\ref{gr}).

An equivalent definition of a pure spinor can be given \cite{chev}, according to which 
all $\eta$-bilinears vanish up to dimension $n=D/2$:
\eq{\label{ch}\tilde{\eta}\gamma_{m_1\dots m_p}\eta=0~;~~~0\leq p<n~,}
where $\tilde{\eta}:=\eta^{\mathrm{Tr}}C^{-1}$, with $C$ the charge-conjugation matrix. 
The insertion of $C^{-1}$ in the above bilinear ensures that the left-hand side above transforms as an (antisymmetric) tensor, cf.~appendix \ref{sec:spinors}.

\color{mygreen}

Let us show that (\ref{ch}) is indeed equivalent to our previous definition of a pure spinor in the simplest case of two Euclidean dimensions, i.e. $n=1$. In a basis where $\gamma_1$ is real  symmetric and $\gamma_2$ is imaginary antisymmetric, there is one creation and 
one annihilation operator which obey, 
\eq{\label{3}(\gamma^{\pm})^{\mathrm{Tr}}=\gamma^{\mp}~.}
Moreover the charge conjugation matrix can be taken to be $C=\gamma_2$, so that
from (\ref{1}) we have,
\eq{C=C^{-1}=i(\gamma^--\gamma^+)~.}
On the other hand $\eta$ obeys, 
\eq{\gamma^-\eta=0=\eta^{\mathrm{Tr}}\gamma^+~,}
where we took (\ref{3}) into account. It follows that 
\eq{\tilde{\eta}\eta={\eta}^{\mathrm{Tr}}C^{-1}\eta=i\eta^{\mathrm{Tr}}(\gamma^--\gamma^+)\eta=0~,}
i.e.~(\ref{ch}) is satisfied. 

The converse is also true, in a trivial way, since in $D=2$ it 
can be shown that every Weyl spinor is pure. This can be seen 
by expanding as in (\ref{gr}),
\eq{\psi_+ = c\eta~;~~~\psi_-=c'\gamma^+\eta~,}
for some $c,c'\in\mathbb{C}$, where $\psi_{\pm}$ are arbitrary Weyl spinors of positive, negative chiralities respectively. Since $\psi_{\pm}$ are annihilated by $\gamma^{\mp}$ respectively, they are both pure spinors.

It can similarly be shown that all Weyl spinors are pure, up to and including $D=6$. In dimensions $D\geq 8$ not every Weyl spinor is pure.

\color{black}

An important implication of the above is that 
given a Riemannian spin manifold $(\mcal,g)$, a 
non-vanishing pure spinor $\eta$ on $\mcal$ defines an almost complex structure on $\mcal$, i.e. a spliting of the tangent space into holomorphic and antiholomorphic directions, 
\eq{T_s^+\mathcal{M}\oplus T_s^-\mathcal{M}= T^{\mathbb{C}}_s\mathcal{M}~,}
at each point $s\in\mcal$. 
Indeed given $\eta$, equation (\ref{ps}) can be thought of as selecting the   antiholomorphic gamma matrices. Equivalently we may view (\ref{ps}) as defining 
which vectors are holomorphic:
\eq{\label{z}
 v\in T^+\mathcal{M}
\xLeftrightarrow{\mathrm{def}} v^i\gamma_i\eta=0~.}
I.e. $v$ is holomorphic if its  spinorial action annihilates $\eta$, see below (\ref{clal}). These relations do not change by rescaling $\eta$, so in fact the correspondence is between line bundles of pure spinors on $\mcal$ and almost complex structures on $\mcal$. The almost complex structure can be constructed explictly as 
a spinor bilinear:
\eq{\label{cs}{I}_i{}^j=-i\eta^\dagger\gamma_i{}^j\eta~;~~~
{I}_i{}^k{I}_k{}^j=-\delta_i^j~.}
We will also introduce a real two-form and a bivector, both denoted by $J$,
\eq{\label{cs1}J_{ij}:={I}_i{}^pg_{pj}~;~~~
J^{ij}:=g^{ip}{I}_p{}^j
~,}
obtained from $I$ by appropriately raising, lowering indices with the metric of $\mcal$. The antisymmetry of $J$ immediately follows from  (\ref{cs}). For later use let us also define,
\eq{\label{cs2}{I}^i{}_j:=g^{ip}{I}_p{}^qg_{qj}=J^{iq}g_{qj}=-J^{qi}g_{qj}=-{I}_i{}^j
~,}
and note that the metric is hermitian with respect to $I$,
\eq{\label{22}{I}_i{}^p{I}_j{}^qg_{pq}=g_{ij}~.}

\noindent\color{mygreen}Let us show the second equation in (\ref{cs}) in $D=6$ for concreteness. Start with the Firez identity in six dimensions,
\eq{\chi\psi^{\dagger}=\sum_{p=0}^6\frac{1}{p!}\big(\psi^{\dagger}\gamma_{m_p\dots m_1}\chi\big)
\gamma^{m_1\dots m_p}~.}
Applying this in the case $\chi,\psi\rightarrow\eta$ and taking the Hodge-duality of gamma matrices into account,
\eq{i\gamma^{(k)}=(-1)^{\frac12 k(k-1)}\star\gamma^{(6-k)}\gamma_7~,}
leads to: 
\eq{\label{f2}\eta\eta^{\dagger}=\frac14 P_++\frac{i}{8}I_{i}{}^{j}\gamma_{j}{}^{i}P_+~,}
where $P_+:=\frac12 (1+\gamma_7)$ is the positive-chirality projector,  and we used the first of (\ref{cs}) to express the $\eta$-bilinear in terms of $I$. Moreover we have 
normalized $\eta^{\dagger}\eta=1$, and we have taken into account that $(\eta^{\dagger}\gamma^{(p)}\eta)$ vanishes for $p$ odd, as can be seen from (\ref{bilins}) and the fact that $\eta$, $\eta^c$ have opposite chiralities in 
$D=6$. We thus obtain:
\eq{\spl{\label{24}
I_a{}^b I_b{}^c&=-(\eta^\dagger\gamma_a{}^b\!\!\!\!\!\!\!\!
\!\!\!\!\underbrace{\eta)(\eta^\dagger}_{\frac14 P_++\frac{i}{8}I_{i}{}^{j}\gamma_{j}{}^{i}P_+}\!\!\!\!\!\!
\!\!\!\!\!\!\gamma_b{}^c\eta)\\
&=-\frac54\delta_a^c\underbrace{(\eta^\dagger\eta)}_{1}
+iI_a{}^c-\underbrace{(\eta^\dagger\gamma_a{}^c\eta)}_{iI_a{}^c}
+\frac{i}{2}I_a{}^b (\eta^\dagger\gamma_b{}^c\eta)
+\frac{i}{2}I_b{}^c (\eta^\dagger\gamma_a{}^b\eta)
+\frac{i}{8}\delta_a^cI_d{}^e (\eta^\dagger\gamma_e{}^d\eta)\\
&=-\frac54\delta_a^c-I_a{}^b I_b{}^c 
+\frac18\delta_a^cI_d{}^e I_e{}^d
~,}}
where to go from the first to the second line we used the Fierz identity (\ref{f2}) and we expanded out the resulting products of gamma matrices. 
Contracting (\ref{24}) with $\delta_c^a$ gives $I_d{}^e I_e{}^d=-6$, and 
the second equation in (\ref{cs}) then follows by plugging this result back into (\ref{24}).

\color{black}

\subsection{Supergravity backgrounds}

We will consider supergravity backgrounds where spacetime splits onto an internal and an external part, which for our purposes we may take to be flat Minkowski space. The internal 
space will be taken to be a Riemannian spin manifold of even dimension  $(\mathcal{M}_{2n},g)$. 
We are interested in bosonic supersymmetric vacua (solutions) of type II supergravity, i.e. all fermionic supergravity fields will be set to 
zero at the solution. Moreover we will consider the case where the supersymmetry 
of the vacuum is parameterized by (pairs) of nowhere-vanishing pure spinors on $\mathcal{M}_{2n}$.\footnote{In the present setup the nowhere-vanishing property of the spinor parameters is simply a consequence of the Killing spinor equations.} We will call these {\it pure backgrounds}. 

\vfil\break

Recall that an $SU(n)$ structure on $\mathcal{M}_{2n}$ is given by,
\begin{itemize}
\item a complex decomposable $n$-form  $\Omega$,
\item a real two-form $J$ such that,
\item $\Omega\wedge J=0$ ~~~\& ~~$\displaystyle\frac{i^{n(n+2)}}{2^n}\Omega\wedge\Omega^* = \frac{1}{n!}J^n$~.
\end{itemize}
We then have the following 
equivalences:
\begin{enumerate}
\item Reduction of the structure group of $\mcal_{2n}$ to $SU(n)$
\item  Existence on  $\mcal_{2n}$ of  $(g$, ${I})$  with $c_1(I)=0$
\item Existence on  $\mcal_{2n}$ of $(g$, $\eta)$ with $\eta$ pure \& nowhere-vanishing
\end{enumerate}

\color{mygreen}

The proof of the equivalence of (2) and (3) follows directly from our discussion 
on the relation between pure spinors and almost complex structures:  
the vanishing of the first Chern class of the almost complex structure $I$ 
is equivalent to the fact that $\eta$ is globally defined and nowhere-vanishing. Moreover 
the fact that (3) implies (1) can be shown  with manipulations similar to those used in the proof 
of (\ref{cs}) given above: besides the spinor bilinear defining $I_m{}^n$ and $J_{mn}=I_m{}^pg_{pn}$, one defines the complex $n$-form via $\Omega=\tilde{\eta}\gamma_{(n)}\eta$. Then by Fierzing one shows the algebraic compatibility conditions between $J$ and $\Omega$ listed above. Finally, to see the equivalence of  (1) and (2) first note that a metric and an almost complex structure define  a reduction of the structure group to $U(n)$. This can be seen by 
using the almost complex structure $I$ to 
put the metric $g$ in a canonical form whose  stabilizer is manifestly given by $U(n)$,
\eq{g=\sum_{a=1}^ne_a\otimes \bar{e}_a~,}
where the $e_a$'s are holomorphic one-forms with respect to $I$. Now consider the holomorphic top form,  
\eq{\label{27}\Omega:=e_1\wedge\dots\wedge e_n~,}
which transforms as a section of the canonical bundle $K$ of $I$. 
Let $t_{(ij)}\in U(n)$ be the transition functions of the holomorphic tangent bundle $T^+\mathcal{M}$ on the overlap of two open patches $U_i$, $U_j\in \mathcal{M}$. This means that if $e_a|_i$, $e_a|_j$ are  the values of the holomorphic one-forms on $U_i$, $U_j$ then,
\eq{\label{28}e_a|_i=\sum_{b=1}^n[t_{(ij)}]_{ab}e_b|_j~.}
From (\ref{27}),(\ref{28}) it follows that,
\eq{\Omega|_i=\det[t_{(ij)}]\Omega|_j}
I.e. $\Omega$ is a section of a  complex line bundle ($K$) with transition functions given by the determinant of the $U(n)$ transition functions of $T^+\mathcal{M}$. It follows that $\Omega$ is globally defined iff $K$ has a global section, which is equivalent to saying that $K$ is trivial ($c_1(I)=0$) and thus $\det[t_{(ij)}]=1$. In other words the transition functions of $T^+\mathcal{M}$ are in $SU(n)$.

\color{black}

The link with type II supergravity can be made by expressing the  
ten-dimensional supersymmetry parameters schematically as follows,
\eq{\epsilon_1~{\sim}~\zeta\otimes {\eta_1}~;~~~ 
\epsilon_2~{\sim}~\zeta\otimes {\eta_2}~,}
where $\eta_a$, $a=1,2$, are taken to be unimodular pure spinors on $\mathcal{M}_{2n}$, 
\eq{\label{unim}
\eta_a^{\dagger}\eta_a=1
~.}
The $\zeta$'s are constant spinors of the external Minskowski space.  
These then define an $SU(n)\times SU(n)$ structure on  $\mathcal{M}_{2n}$ which can be explictly constructed using spinor bilinears: 
\eq{\label{bj}\Omega_a=\tilde{\eta}_a\gamma_{(n)}\eta_a~~~;~ J=-i\eta^{\dagger}_a\gamma_{(2)}\eta_a~,}
for $a=1,2$. The supersymmetry equations of the vacuum (the Killing spinor equations) schematically take the form,
\eq{
\nabla\eta+F\cdot\gamma\eta=0
~,}
where $F\cdot\gamma$ is a Clifford-algebra element determined by the flux. These 
first-order equations then generically impose constraints on the torsion classes of the  $SU(n)\times SU(n)$ structure. 
The upshot is that, 
\begin{center}\vskip -.3cm
\framebox{\bf pure backgrounds  \boldmath$=SU(n)\times SU(n)$\unboldmath -structure backgrounds}
\end{center}

\subsection{Generalized complex geometry}

In generalized geometry one is interested in the generalized tangent bundle, i.e. the sum of tangent and cotangent bundles of $\mathcal{M}$, 
$T\oplus T^{*}$. Consider a generalized vector $V$, 
\eq{V=(a^i\partial_i,b_j\d x^j)\in T\oplus T^{*}~,}
$i,j=1,\dots,D$,  
or in components: 
\eq{V^M=(a^i,b_j)~.}
Note that the generalized index $M$ decomposes into a pair of oridinary tangent and cotangent indices. 
There is a natural action of $V$ on polyforms $\varphi\in\Lambda^\bullet T^*$, given by:
\eq{\label{ga}V^M\Gamma_M\cdot\varphi=(\iota_a+b\wedge )\varphi~,}
which provides a representation of the Clifford algebra $Cl(T\oplus T^{*})$~,
\eq{\label{genga}\{ \Gamma_M, \Gamma_N\}= \mathcal{G}_{MN}~,}
where  
\eq{\label{tr}
\Gamma^M=(\d x^i\wedge,\iota_{\partial_j})
~;~~~
\Gamma_M=\mathcal{G}_{MN}\Gamma^N=(\iota_{\partial_i},\d x^j\wedge)
~,}
and 
\eq{\label{genme}\mathcal{G}=
 \left(\begin{array}{cc} 0&\mathbb{I}_D \\ \mathbb{I}_D & 0 \end{array}\right)~,}
with $\mathbb{I}_D$ the identity $D\times D$ matrix, 
is the indefinite metric with signature $(D,D)$ induced by  the natural  
pairing of vectors and forms,
\eq{\label{norm}\langle V,V\rangle=\sum_{i=1}^Da^i b_i~.}
In other words, comparing with our earlier discussion of the Clifford algebra (\ref{clal}), equation (\ref{ga}) defines the spinorial action of $V$, 
$\rho(V)=V^M\Gamma_M$, where $\Gamma_M$ given in (\ref{tr}) acts on polyforms. 
This representation is faithful and has dimension $2^D$, i.e. it is the least-dimensional representation of the Clifford algebra associated 
with the vector space $T\oplus T^{*}$. We can therefore identify the space of polyforms $\Lambda^\bullet T^*$ with 
the spinor module of $Cl(T\oplus T^{*})$.

\color{mygreen}
It is straightforward to derive the explicit form of $\mathfrak{so}(D,D)$ transformations, i.e. transformations $V^M\rightarrow V^{\prime M}:=R^M{}_NV^N$ that leave the norm (\ref{norm})  invariant. Parameterizing,
\eq{\label{40}
R^M{}_N=\left(\begin{array}{cc}
A^i{}_k & \beta^{il}\\
B_{jk} & Q_j{}^l
\end{array}\right)
~,}
and imposing $\langle V',W\rangle+\langle V,W'\rangle=0$, for all $V,W\in T\oplus T^*$, can be seen, after a little bit of algebra, to be equivalent to the requirement that $B_{ij}$, $\beta^{ij}$ are antisymmetric and $Q_i{}^j=-A^j{}_i$. In other words generic $\mathfrak{so}(D,D)$ transformations decompose into $\mathfrak{gl}(D)$ transformations parameterized by $A$ and the so-called $B$- and $\beta$-transforms parameterized by the two-form $B$ and the bivector $\beta$  respectively.

Using the explicit form of $R^M{}_N$  derived above, it is also straightforward to compute the spinorial action of $\mathfrak{so}(D,D)$ transformations on polyforms $\varphi$ (i.e. Dirac spinors of $T\oplus T^*$). Explicitly we have:
\eq{\varphi\rightarrow\varphi'=\frac12 R_{MN}\Gamma^{MN}\varphi~,}
where we use the metric (\ref{genme}) to raise, lower generalized indices. 
Taking the definition of $\Gamma_M$ into account and inserting above leads to,
\eq{\spl{\label{244}\varphi\rightarrow\varphi'
&=\frac12 \big\{
A^i{}_j\iota_{\partial_i}\d x^j\wedge+
\underbrace{Q_j{}^i}_{-A^i{}_j}\d x^j\wedge\iota_{\partial_i}
+\beta^{ij}\iota_{\partial_i}\iota_{\partial_j}
+B_{ij}\d x^i\wedge \d x^j\wedge
\big\}\varphi
\\
&=\frac12 \left\{
A^i{}_j[\iota_{\partial_i},\d x^j\wedge]+\beta^{ij}\iota_{\partial_i}\iota_{\partial_j}
+B_{ij}\d x^i\wedge \d x^j\wedge
\right\}\varphi~.}}
We see that $B$-transforms act by wedging with $B$, $\varphi\rightarrow
 B\wedge\varphi$, while $\beta$-transforms act by contraction with $\beta$, $\varphi\rightarrow
 \iota_{\beta}\varphi$. Furthermore the $\mathfrak{gl}(D)$ transformations 
parameterized by $A$ give,
\eq{\varphi\rightarrow \frac12 
A^i{}_j\iota_{\partial_i}(\d x^j\wedge\varphi)-\frac12 
A^i{}_j\d x^j\wedge\iota_{\partial_i}\varphi 
=\frac12 A^i{}_i\varphi- 
A^i{}_j\d x^j\wedge\iota_{\partial_i}\varphi
~.}
We recognize the second term on the right-hand side as the standard action of $\mathfrak{gl}(D)$ on forms. Exponentiating the result above we thus obtain the action of $GL(D)$ transformations on polyforms induced by the spinorial action of generalized vectors,
\eq{\label{44}M\cdot\varphi=\sqrt{\det M}M_*\varphi~,}
where $M=\exp A\in GL(D)$ and $M_*$ is the standard action of $M$ on forms.

\color{black}

On the other hand, under $SO(D)$ structure group transformations of the base manifold $\mcal$,  polyforms can be identified with bispinors. This is explicitly realized by Fierzing and the so-called {\it Clifford map}:
\eq{\spl{\label{fbi}\psi_\alpha\otimes\widetilde{\chi}_\beta~
=
~&\frac{1}{2^n}\sum_{p=0}^{2n}\frac{1}{p!}
(\widetilde{\chi}\gamma_{m_p\dots m_1}\psi)~\!\gamma^{m_1\dots m_p}_{\alpha\beta}\\
{\leftrightarrow}
~&\frac{1}{2^n}\sum_{p=0}^{2n}\frac{1}{p!}
(\widetilde{\chi}\gamma_{m_p\dots m_1}\psi)~\!\d x^{m_1}\wedge\dots\wedge \d x^{m_p}~,}}
which identifies antisymmetric products of gamma matrices with forms.

\color{mygreen}

Let us show that the two sides of the identification (\ref{fbi}) transform in the same way 
under $SO(D)$ transformations. 
Let $R_{ij}=R_{[ij]}$ be an element of the Lie algebra of  $\mathfrak{so}(D)$. The corresponding spinor transformation reads,
\eq{\delta_R\psi=\frac14 R_{ij}\gamma^{ij}\psi
~;
~~~\delta_R\widetilde{\chi}=
-\frac14 R_{ij}\widetilde{\chi}\gamma^{ij}
~.}
We thus obtain,
\eq{\spl{\label{re}\delta_R (\widetilde{\chi}\gamma_{m_1\dots m_p}\psi)&= 
\frac14 {R}_{ij}(\widetilde{\chi}\big[\gamma_{m_1\dots m_p},\gamma^{ij}\big]\psi)\\
&=p{R}_{[m_p}{}^{i} (\widetilde{\chi}\gamma_{m_1\dots m_{p-1}]i}\psi)
~,}}
where we recognize the  right hand side as the standard action of $\mathfrak{so}(D)$ on forms. Exponentiating (\ref{re}) we obtain the action $M\cdot$ on a polyform $\varphi$ of an element $M\in SO(D)$ induced by the action of $M$ on spinors via the map (\ref{fbi}),
\eq{M\cdot\varphi= M_*\varphi~,}
where $M_*$ denotes the standard action of $M$ on forms. Note also that the formula above can be thought of as a specialization of (\ref{44}) for $SO(D)\subset GL(D)$.

\color{black}

To summarize, 
\begin{center}\vskip -.3cm
\framebox{\bf polyforms on \boldmath$\mcal=$\unboldmath {} spinors of \boldmath$Cl(T\oplus T^{*})$\unboldmath {} \boldmath$=$\unboldmath {} bispinors on \boldmath$\mcal$\unboldmath}
\end{center}

It is straightforward, under the identification (\ref{fbi}), to read off how the 
left and right action of ordinary gamma matrices on bispinors translates to 
an action on polyforms. We find:
\eq{\spl{\label{21}
\gamma_m\underline{\Psi}~
&{\leftrightarrow}~
(\iota_{\partial_m}+g_{mk}\d x^k\wedge)~\!\Psi\\
\underline{\Psi}~\!\gamma_m~
&{\leftrightarrow}~
(-\iota_{\partial_m}+g_{mk}\d x^k\wedge)~\!\Psi~\!(-1)^{|\Psi|}
~,}}
equivalently,
\eq{\spl{\label{p}\iota_{\partial_m}\Psi&\leftrightarrow
\frac12 \big(\gamma_m\underline{\Psi}-(-1)^{|\Psi|}\underline{\Psi}\gamma_m\big)\\
dx^m\wedge\Psi&\leftrightarrow
\frac12 \big(\gamma^m\underline{\Psi}+(-1)^{|\Psi|}\underline{\Psi}\gamma^m\big)
~,}}
where an underline denotes the image of a polyform under the Clifford map, i.e. the corresponding bispinor: 
\eq{\label{cm}
\underline{\Psi}:=\sum_{p=1}^D\frac{1}{p!}\Psi_{m_1\dots m_p}\gamma^{m_1\dots m_p}\leftrightarrow 
\sum_{p=1}^D\frac{1}{p!}\Psi_{m_1\dots m_p} dx^{m_1} \wedge\dots \wedge dx^{m_p}=\Psi~.}
The sign $(-1)^{|\Psi|}$ in the second line of (\ref{21}) is defined to be positive, negative for 
even, odd polyforms respectively. Moreover it is equal to the {\it chirality} of the polyform, thought of as a generalized spinor  of $T\oplus T^*$.

\color{mygreen}

To show eqn.~(\ref{21}), or (\ref{p}), note that from  
(\ref{cm}) we obtain,
\eq{\spl{
\gamma_m\underline{\Psi}
&=
~\frac{1}{2^n}\sum_{p=0}^{2n}\frac{1}{p!}
\Psi_{m_1\dots m_p}\!\!\!\!\!\!\!\!\!\!\!\!\!\!
\underbrace{\gamma_m\gamma^{m_1\dots m_p}}_{p\delta_m^{[m_1}\gamma^{m_2\dots m_p]}+\gamma_m{}^{m_1\dots m_p}}\\
~&\leftrightarrow 
\frac{1}{2^n}\sum_{p=0}^{2n}\frac{1}{{(p-1)!}}
\Psi_{mm_2\dots m_p}~\!dx^{m_2}\wedge\dots\wedge dx^{m_p}\\
&~~~~ +
\frac{1}{2^n}\sum_{p=0}^{2n}\frac{1}{p!}
\Psi_{m_1\dots m_p}~\!g_{mk}dx^{k}\wedge dx^{m_1}\wedge\dots\wedge dx^{m_p}\\
&=
\frac{1}{2^n}\sum_{p=0}^{2n}\frac{1}{p!}
\Psi_{m_1\dots m_p}~\!(\iota_m+g_{mk}\d x^k\wedge)\big(dx^{m_1}\wedge\dots\wedge dx^{m_p}\big)\\
&=(\iota_m+g_{mk}\d x^k\wedge)~\!\Psi
~.}}
The second line of (\ref{21}) can be established in a similar manner.

Let us also show that even, odd forms have positive, negative chirality  respectively, when thought of as generalized spinors. 
First it will be useful to pass to a more conventional basis of 
generalized gamma matrices, diagonalizing the metric in (\ref{genga}). Explicitly,
\eq{
\{\hat{\Gamma}_M,\hat{\Gamma}_N\}=2\mathcal{H}_{MN}~;~~~T\cdot\mathcal{G}\cdot T=2\mathcal{H}~;~~~\hat{\Gamma}:=T\cdot\Gamma
~, }
where,
\eq{\mathcal{H}:=\left(\begin{array}{cc}
-\mathbb{I}_D&0\\
0&\mathbb{I}_D
\end{array}\right)~;~~~
T:=\left(\begin{array}{cc}
-\mathbb{J}_D&\mathbb{J}_D\\
\mathbb{J}_D&\mathbb{J}_D
\end{array}\right)
~,}
and $\mathbb{J}_D$ is the $D\times D$ matrix with units along the 
NE-SW diagonal and zeros everywhere else. The generalized gamma matrices 
in the transformed basis read,
\eq{\hat{\Gamma}_M=(-\iota_{\partial_D}+dx^D\wedge,\dots, 
-\iota_{\partial_1}
+dx^1\wedge, \iota_{\partial_D}+dx^D\wedge, \dots, \iota_{\partial_1}+dx^1\wedge)~,}
where we took (\ref{tr}) into account. 
Suppose now that $\Phi$ is an even or odd polyform. 
It is straightforward to compute the action of the $2D$-dimensional chirality matrix $\Gamma_{2D+1}$ on $\Phi$,
\eq{\spl{
\Gamma_{2D+1}~\!\Phi&=i^D\hat{\Gamma}_1\dots \hat{\Gamma}_{D}\hat{\Gamma}_{D+1}\dots\hat{\Gamma}_{2D}~\!\Phi\\
&=i^D(-\iota_{\partial_D}+dx^D\wedge)\dots\!\!\!\!\!\!\underbrace{(-\iota_{\partial_1}
+dx^1\wedge)}_{
(-1)^{|\Phi|}\gamma_1~\mathrm{on~the~right}
}\!\!\!\!
(\iota_{\partial_D}+dx^D\wedge)\dots\underbrace{(\iota_{\partial_1}+dx^1\wedge)}_{
\gamma_1~\mathrm{on~the~left}
}
\Phi\\
&\leftrightarrow i^D\!\!\!\!\!\!\!\!\!\!\!\!\!\!\!\underbrace{\gamma_D\dots\gamma_1}_{
(-i)^n\gamma_{D+1}(-1)^{\frac12 D(D-1)}}\!\!\!\!\!\!\!\!\!\!\!\!\!
\underline{\Phi}~\underbrace{(-1)^{|\Phi|}\gamma_1\dots(-1)^{
|\Phi|}\gamma_D}_{(-i)^n\gamma_{D+1}}\\
&=\gamma_{D+1}\!\!\!\!\!\!\underbrace{\underline{\Phi}~\!\gamma_{D+1}}_{
(-1)^{|\Phi|}\gamma_{D+1}\underline{\Phi}}\\
&=(-1)^{|\Phi|}\underline{\Phi}~,
}}
where we made use of (\ref{21}),(\ref{cm}) and the fact that $\g_{(p)}\g_{D+1}=(-1)^{p}\g_{D+1}\g_{(p)}$.

\color{black}

Coming back to the pure supergravity background parameterized 
by $\eta_{1,2}$, it follows immediately that the generalized spinors, 
\eq{\label{ff}
\underline{\Psi}_1~=~\eta_1\otimes\widetilde{\eta^c_2}
~;~~~\underline{\Psi}_2~=~\eta_1\otimes\widetilde{\eta}_2
~,}
are pure, since they are annihilated by half of the generalized 
gamma matrices. Let us denote by $\mathcal{I}_{1,2}$ the generalized 
almost complex structures (GACS) associated with $\Psi_{1,2}$. Explicitly we have,
\eq{\spl{\label{ee}
\mathcal{I}_{1M}{}^N &= \frac{1}{2}
\left(\begin{array}{cc} I_{1i}{}^j-I_{2i}{}^j & J_{1il}+J_{2il}\\ 
J_1{}^{jk}+J_2{}^{jk}& I_{1}{}^j{}_l-I_{2}{}^j{}_l \end{array}\right) \\
\mathcal{I}_{2M}{}^N &= \frac{1}{2}
\left(\begin{array}{cc} I_{1i}{}^j+I_{2i}{}^j & J_{1il}-J_{2il}\\ 
J_1{}^{jk}-J_2{}^{jk}& I_{1}{}^j{}_l+I_{2}{}^j{}_l \end{array}\right)
~,}}
where ${I}_{1,2}$ are the almost complex structures associated with $\eta_{1,2}$, and we took (\ref{cs1}),(\ref{cs2}) into account. Moreover  it can easily be checked that the GACS commute, and that $\mathcal{G}$ is hermitian with respect to both of them: 
\eq{\label{f}[\mathcal{I}_{1},\mathcal{I}_{2}]=0~;~~~ \mathcal{I}_{aM}{}^P\mathcal{I}_{aN}{}^Q\mathcal{G}_{PQ}=\mathcal{G}_{MN}~,} 
for $a=1,2$. 
Furthermore, the metric of the base manifold $\mcal$ sits inside the product of the two GACS,
\eq{\label{g}\mathcal{Q}_M{}^N:=-\mathcal{I}_{1M}{}^P\mathcal{I}_{2P}{}^N
\xlongequal{[\mathcal{I}_1,\mathcal{I}_2]=0}
-\mathcal{I}_{2M}{}^P\mathcal{I}_{1P}{}^N
={\left(\begin{array}{cc} 0 & g_{il}\\ 
g^{jk}& 0 \end{array}\right)}~.}
We will come back to the properties of $\mathcal{Q}$ in the following, cf. below (\ref{r2}).

\color{mygreen}

To see that  $\Psi_1$ is pure, note that it is annihilated by  half of generalized gamma matrices,
\eq{\label{a}L_1^{\pm}\Psi_1=0 ~,}
where 
$L_{1}^{+}$ represents the left action of $\{\gamma_p^{(1-)},p=1,\dots,D\}$, i.e. the space of (ordinary) gamma matrices 
antiholomorphic with respect to the almost complex structure associated with $\eta_1$. Similarly 
$L_{1}^{-}$ represents the right action of $\{\gamma_p^{(2+)},p=1,\dots,D\}$, i.e. the space of gamma matrices holomorphic with respect to the almost complex structure associated with $\eta_2$. 
Moreover $\Psi_2$ is annihilated by half of the generalized 
gamma matrices,
\eq{\label{b}L_2^{\pm}\Psi_2=0 ~,}
where 
$L_{2}^{+}=L_{1}^{+}$,  while 
$L_{2}^{-}$ represents the right action of $\{\gamma_p^{(2-)},p=1,\dots,D\}$, i.e. the space of gamma matrices antiholomorphic with respect to the almost complex structure associated with $\eta_2$. 

Using the polyform-bispinor correspondence and taking (\ref{21}) into account, the annihilators $L$ can be 
represented on polyforms as follows,
\eq{\spl{\label{25}
L_{1,2}^{(+)}&=\left[\Pi_1^{-}\right]_p{}^m (\iota_m+g_{mk}\d x^k\wedge)\\
L_{1}^{(-)}&=
\left[\Pi_2^{+}\right]_p{}^m (-\iota_m+g_{mk}\d x^k\wedge)\\
L_{2}^{(-)}&=
\left[\Pi_2^{-}\right]_p{}^m (-\iota_m+g_{mk}\d x^k\wedge)~;~~~p=1,\dots,D
~,}}
where $\Pi_{1,2}^{\pm}$ are holomorphic, antiholomorphic projectors with respect to $I_{1,2}$ respectively. Explicitly,
\eq{\Pi_a^{\pm}:=\frac12 \left(\mathbb{I}_D\mp i I_a\right)~;~~~a=1,2~.}
Let us denote by $v_{1}^{(\pm)}$ a vector which is holomorphic, antiholomorphic with respect to the almost complex structure associated with $\eta_1$, and similarly for $v_{2}^{(\pm)}$. Using (\ref{25}), eqs.~(\ref{a}),(\ref{b}) can then be written equivalently as,
\eq{\spl{
&v_{1}^{(+)m}(\iota_m+g_{mk}\d x^k\wedge)\Psi_1=0~;~~~
v_{2}^{(-)m} (-\iota_m+g_{mk}\d x^k\wedge)\Psi_1=0\\
&v_{1}^{(+)m}(\iota_m+g_{mk}\d x^k\wedge)\Psi_2=0~;~~~
v_{2}^{(+)m} (-\iota_m+g_{mk}\d x^k\wedge)\Psi_2=0~;~~~\forall~v_{1}^{(+)},~
v_{2}^{(\pm)}
~,}}
where we took into account that $v_{a}^{(\pm)m}=v_{a}^{(\pm)p}\left[\Pi_a^{\mp}\right]_p{}^m$, for $a=1,2$.
In other words the generalized pure spinor $\Psi_1$ is annihilated 
by the spinorial action of the generalized vectors $V^{(1)}$, $\tilde{V}^{(1)}$, 
\eq{
V_{M}^{(1)}\Gamma^M\Psi_1=\tilde{V}_{M}^{(1)}\Gamma^M\Psi_1=0~,
}
where 
\eq{
V_{M}^{(1)}:=\big(v^{(1+)}_i,v^{(1+)j}\big)~;~~~
\tilde{V}_{M}^{(1)}:=\big(v^{(2-)}_i,-v^{(2-)j}\big)
~.}
Similarly, the generalized pure spinor $\Psi_2$ is annihilated 
by the spinorial action of the generalized vectors $V^{(2)}$, $\tilde{V}^{(2)}$, 
\eq{
V_{M}^{(2)}\Gamma^M\Psi_2=\tilde{V}_{M}^{(2)}\Gamma^M\Psi_2=0~,
}
where
\eq{
V_{M}^{(2)}:=\big(v^{(1+)}_i,v^{(1+)j}\big)~;~~~
\tilde{V}_{M}^{(2)}:=\big(v^{(2+)}_i,-v^{(2+)j}\big)
~.}
Recall that in the ordinary case a pure spinor $\eta$ is annihilated by the 
spinorial action of vectors which are homomorphic with respect to the 
almost complex structure associated with $\eta$, cf. eqn.~(\ref{z}),
\eq{
v_{i}^{(+)}\gamma^i\eta=0~\Leftrightarrow~
I_i{}^jv_{j}^{(+)}=+iv_{i}^{(+)}
~.}
Similarly the generalized pure spinors $\Psi_{1,2}$ are associated with 
the GACS $\mathcal{I}_{1,2}$ whose $+i$-eigenvectors are the $V$, $\tilde{V}$ defined previously,
\eq{
\mathcal{I}_{aM}{}^N V^{(a)}_{N}=+i{V}^{(a)}_{M}~;~~~
\mathcal{I}_{aM}{}^N \tilde{V}^{(a)}_{N}=+i\tilde{V}^{(a)}_{M}
~,}
for $a=1,2$. Knowledge of the eigenvectors 
 then allows us to construct the GACS $\mathcal{I}_{1,2}$ 
explicitly with the result given in (\ref{ee}). Finally, eqs.~(\ref{f}),(\ref{g}) can 
be verified directly using the explicit expression (\ref{ee}).

\color{black}

Alternatively $\mathcal{I}_{a}$, for $a=1,2$, can be expressed as generalized-spinor 
bilinears,
\eq{\label{6}
\mathcal{I}_{aM}{}^N  =-2i
\frac{\langle\Psi_a^{*},\Gamma_{M}{}^N \Psi_a\rangle}{\langle\Psi_a^{*}, \Psi_a\rangle}
~,}
where we have introduced the Mukai pairing of two polyforms $\Phi_1$, $\Phi_2$:
\eq{\label{8}
\langle\Phi_1,\Phi_2\rangle:=\left.\Phi_1\wedge\sigma(\Phi_2)\right|_{D}
~.}
The involution $\sigma$ above inverts the order of form indices,
\eq{\sigma\left(dx^{m_1}\wedge dx^{m_2}\wedge\dots\wedge dx^{m_p}\right):=dx^{m_p}\wedge\dots \wedge dx^{m_2}\wedge dx^{m_1}~,} 
so that for a $p$-form $\varphi$, $\sigma(\varphi)=(-1)^{\frac12 p(p-1)}\varphi$. 
Alternatively the Mukai pairing can be expressed in terms of bispinors,
\eq{\label{7}
\langle\Phi_1,\Phi_2\rangle
\propto
\mathrm{tr}(\underline{\widetilde{\Phi}_1}
\gamma_{D+1}\underline{\Phi_2})~\!
\mathrm{vol}_{D}
~,}
where the proportionality constant depends on the dimension $D$, and
\eq{\label{9}\widetilde{\underline{\Phi}}:=
C\underline{\Phi}^{\mathrm{Tr}}C^{-1}~.}

\color{mygreen}

Let us work in $D=10$ for concreteness; 
the calculation is similar in other dimensions. 
To show the equivalence of (\ref{8}),(\ref{7}), first note that 
definition (\ref{9}) implies,
\eq{\label{78}\widetilde{\gamma}_{m_1\dots m_p}=(-1)^{\frac12 p(p+1)}\gamma_{m_1\dots m_p}~,}
as follows from the gamma-matrix property $\gamma_m^{\mathrm{Tr}}=(-1)^nC^{-1}\gamma_mC$, with $D=2n$. We have,
\eq{\spl{\label{10dm}
\frac{1}{2^5}\mathrm{tr}(\underline{\widetilde{\Phi}}
\gamma_{d+1}\underline{\Psi})~\!
\mathrm{vol}_{10}&=
\sum_{p+q=10}\frac{1}{2^5p!q!}\mathrm{tr}
\big(\!\!\!\!\!\!\!\!\!\!\underbrace{\widetilde{\gamma}^{m_1\dots m_p}\gamma_{11}{\gamma}^{n_1\dots n_q}}_{(-1)^{\frac12 p(p-1)}\gamma_{11}{\gamma}^{m_1\dots m_p}{\gamma}^{n_1\dots n_q}}
\!\!\!\!\!\!\!\!\!\!\big)
\Phi_{m_1\dots m_p}\Psi_{n_1\dots n_q}\mathrm{vol}_{10}\\
&=
\sum_{p+q=10}(-1)^{\frac12 p(p-1)}\frac{1}{2^5p!q!}\underbrace{\mathrm{tr}
\big(
\gamma_{11}{\gamma}^{m_1\dots m_p n_1\dots n_q}
\big)}_{-i2^5\varepsilon^{m_1\dots m_p n_1\dots n_q}}
\Phi_{m_1\dots m_p}\Psi_{n_1\dots n_q}\mathrm{vol}_{10}
\\
&=-i\sum_{p+q=10}(-1)^{\frac12 p(p-1)}\frac{1}{p!q!}\Phi_{m_1\dots m_p}\Psi_{n_1\dots n_q}\!\!\!\!\!\!\!\!\!\!
\underbrace{\varepsilon^{m_1\dots m_p n_1\dots n_q}\mathrm{vol}_{10}}_{
dx^{m_1} \wedge\dots \wedge dx^{m_p}\wedge dx^{n_1}\wedge\dots \wedge dx^{n_q}
}\\
&=-i\sum_{p+q=10}(-1)^{\frac12 p(p-1)}\Phi\wedge\!\!\!\!\!\!\!\!\!\!\!\!\!\!\!\!\!\underbrace{\Psi}_{(-1)^{\frac12 (p-1)(p-2)}\sigma(\Psi)}\\
&=i\sum_{p+q=10}(-1)^{p}\Phi\wedge\sigma(\Psi)\\
&=i(-1)^{|\Phi|}\langle\Phi,\Psi\rangle~.
}}
Equation (\ref{6}) is the generalized-geometry version of the second 
equation in (\ref{bj}) which expresses the almost complex structure as a spinor 
bilinear.  To show (\ref{6}) first consider $(M,N)=(i,n+j)$, with $i,j\leq D$. From (\ref{tr}) we have,
\eq{\spl{\label{11}
\Gamma_M{}^N\Psi=\Gamma_{ij} \Psi&=\iota_{\partial_i}\iota_{\partial_j}\Psi_1\\
&\leftrightarrow \frac{1}{4}\Big[
\gamma_i\big(\gamma_j\underline{\Psi}-(-1)^{|\Psi|}\underline{\Psi}\gamma_j\big)
-(-1)^{|\Psi|+1}
\big(\gamma_j\underline{\Psi}-(-1)^{|\Psi|}\underline{\Psi}\gamma_j\big)\gamma_i
\Big]\\
&= \frac{1}{4}\big(
\gamma_{ij}\underline{\Psi}+\underline{\Psi}\gamma_{ij}
-2(-1)^{|\Psi|}\gamma_{[i}\underline{\Psi}\gamma_{j]}
\big)
~,
}}
where to go from the first to the second line above we used (\ref{p}). 
In the following let us take $\Psi=\Psi_1$,  cf.(\ref{ff}), with 
$\eta_{1,2}$ positive-chirality pure spinors (this setup is 
akin to e.g. Euclidean ten-dimensional IIB supergravity). This implies that $\eta_2^c$ is of negative chirality and therefore 
$\underline{\Psi_1}$ is an even polyform, as follows from \ref{bilins} (similarly it can be seen that $\underline{\Psi_2}$ is an odd polyform). 
Using (\ref{fbi}) we have,
\eq{\spl{\label{above}
\Psi_1&=
\frac{1}{2^5}\sum_{p=\mathrm{even}}\frac{1}{p!}
(\widetilde{\eta_2^c}\gamma_{m_p\dots m_1}\eta_1)~\!dx^{m_1}\wedge\dots\wedge dx^{m_p}\\
\xRightarrow{(\ref{bilinscomp})}\Psi_1^*&=-
\frac{1}{2^5}\sum_{p=\mathrm{even}}\frac{1}{p!}
(\widetilde{\eta_2}\gamma_{m_p\dots m_1}\eta_1^c)~\!dx^{m_1}\wedge\dots\wedge dx^{m_p}\\
\Rightarrow\underline{\Psi_1^*}&=-
\frac{1}{2^5}\sum_{p=\mathrm{even}}\frac{1}{p!}
(\widetilde{\eta_2}\gamma_{m_p\dots m_1}\eta_1^c)~\!\gamma^{m_1\dots m_p}\\
\xRightarrow[(\ref{78})]{(\ref{bilinstr})}\widetilde{\underline{\Psi_1^*}}
&=\frac{1}{2^5}\sum_{p=\mathrm{even}}\frac{1}{p!}
(\widetilde{\eta_1^c}\gamma_{m_p\dots m_1}\eta_2)~\!\gamma^{m_1\dots m_p}\\
&={\eta_2}\otimes\widetilde{\eta_1^c}
~.}}
Pluging (\ref{11}),(\ref{above}) into the expression (\ref{10dm}) of the Mukai pairing in $D=10$, we can calculate 
the generalized spinor bilinear,
\eq{\spl{
\label{76}\langle\Psi_1^{*},\Gamma_M{}^N \Psi_1\rangle
/\mathrm{vol}_{10}
&=-i(-1)^{|\Psi_1|}\frac{1}{2^5} \mathrm{tr}\big(
\widetilde{\underline{\Psi^*_1}}\gamma_{11}\underline{\Gamma_{ij}\Psi_1}
\big)\\
&=-\frac{i}{2^7} \mathrm{tr}\big[
{\eta_2}\otimes\widetilde{\eta_1^c}\gamma_{11}\big(
\gamma_{ij}\underline{\Psi_1}+\underline{\Psi_1}\gamma_{ij}
-2\gamma_{[i}\underline{\Psi_1}\gamma_{j]}
\big)
\big]\\
&=-\frac{i}{2^7}\big[
\widetilde{\eta_1^c}\big(
\gamma_{ij}\underline{\Psi_1}+\underline{\Psi_1}\gamma_{ij}
-2\gamma_{[i}\underline{\Psi_1}\gamma_{j]}
\big){\eta_2}
\big]\\
&=-\frac{i}{2^7}\big[
\underbrace{(\widetilde{\eta_2^c}{\eta_2})}_{-1}
\underbrace{(\widetilde{\eta_1^c}\gamma_{ij}{\eta_1})}_{-iJ_{1ij}}
+\underbrace{(\widetilde{\eta_1^c}{\eta_1})}_{-1}
\underbrace{(\widetilde{\eta_2^c}\gamma_{ij}{\eta_2})}_{-iJ_{2ij}}
-2(\widetilde{\eta_1^c}\gamma_{[i}{\eta_1})(\widetilde{\eta_1^c}\gamma_{j]}{\eta_1})
\big]\\
&=\frac{1}{2^7}(J_{1ij}+J_{2ij})
~,
}}
where we have taken into account that $\Psi_1$ is an even polyform 
and $\eta_1$ has positive chirality. To go from the penultimate to the last line 
we have used (\ref{unim}),(\ref{bj}), taking (\ref{a3}) into account, and noted that $\widetilde{\eta_a^c}\gamma_i\eta_a$ vanishes for $a=1,2$, as
can be seen from the second line of (\ref{bilins}) and the 
fact that ${\eta_a^c}$, $\eta_a$ have opposite chiralities. 
The denominator in (\ref{6}) is calculated similarly to give,
\eq{\langle\Psi_1^{*},\Psi_1\rangle
/\mathrm{vol}_{10}
=-\frac{i}{2^5}
~.
}
Repeating the calculation for different ranges of indices and inserting into (\ref{6}) then leads to (\ref{ee}).

\color{black}

Let us recapitulate: Starting from an $SU(n)\times SU(n)$ background we have constructed a pair of generalized pure spinors and their associated commuting, metric-compatible globally-defined GACS $\mathcal{I}_{1,2}$. 

Conversely, the existence of a pair of globally-defined GACS $\mathcal{I}_{1,2}$ 
obeying (\ref{f}) can be seen to lead to the reduction of the structure group of the generalized tangent bundle to  $SU(n)\times SU(n)$.\footnote{In addition we must assume that the generalized metric defined in (\ref{gm}) below is positive definite.} The most general solution for  $\mathcal{I}_{1,2}$ can be seen to encode a metric and a 
$B$-field on $\mcal$. Explicitly, the most general GACS are 
obtained from (\ref{ee}) by a $B$-transform,
\eq{\label{r1}\mathcal{I}_a\rightarrow \exp(R)\cdot\mathcal{I}_a\cdot\exp(-R)~,
}
for $a=1,2$, where,
\eq{\label{r}
R:=\left(\begin{array}{cc} 0&0 \\ B & 0 \end{array}\right)
~,}
is an element of the Lie algebra $\mathfrak{so}(D,D)$, and $B$ is a two-form. 
Correspondingly the most general generalized pure spinors $\Psi_a$ associated  with $\mathcal{I}_a$ are 
obtained from (\ref{ff}) by the spinorial action of the $B$-transform (\ref{r}),
\eq{\label{r2}
\Psi_a\rightarrow \exp\big(\frac{1}{2}R_{MN}\Gamma^{MN}\big)\Psi_a=e^B\wedge\Psi_a
~.}

\color{mygreen}

To show (\ref{r1}),(\ref{r2}) first note that 
$\mathcal{Q}_M{}^N$ defined in (\ref{g}) obeys, 
\eq{\label{p1}\mathcal{Q}_M{}^P\mathcal{Q}_P{}^N=\delta_M^N~,} 
as follows from its definition and the fact that the two generalized almost complex structures commute, cf.~(\ref{f}). Moreover, 
lowering one index with the canonical 
metric (\ref{genme}), defines what is sometimes refered to as the {\it generalized metric},
\eq{\spl{\label{gm}\mathcal{Q}_{MN}&:=\mathcal{Q}_M{}^S\mathcal{G}_{SN}\\
&=-I_{1M}{}^{R}\underbrace{I_{2R}{}^{S}\mathcal{G}_{SN}}_{
-I_{2N}{}^{S}\mathcal{G}_{SR}
}\\
&=I_{1M}{}^{R}I_{2N}{}^{S}\mathcal{G}_{RS}
~,}}
where to go from the second to the last line we used the hermiticity of the 
canonical metric, cf.~(\ref{f}), which implies in particular $I_{2(N}{}^{S}\mathcal{G}_{R)S}=0$.  
A similar manupilation gives,
\eq{\mathcal{Q}_{MN}
=-I_{2M}{}^{R}\underbrace{I_{1R}{}^{S}\mathcal{G}_{SN}}_{
-I_{1N}{}^{S}\mathcal{G}_{SR}
}
=I_{2M}{}^{R}I_{1N}{}^{S}\mathcal{G}_{RS}
~.}
Comparing the two equations above we obtain,
\eq{\label{p2} \mathcal{Q}_{MN}=\mathcal{Q}_{NM}
~.}
Now suppose there is a $\mathcal{Q}'$ obeying (\ref{p1}), (\ref{p2}). A little bit of matrix algebra shows that the most general 
solution is of the form,
\eq{\label{86}\mathcal{Q}'^M{}_N= 
\left(\begin{array}{cc} 1 & 0\\ 
B &  1 \end{array}\right)\cdot
\left(\begin{array}{cc} 0 & ~~\!g^{-1}\\ 
g & 0 \end{array}\right)
\cdot
\left(\begin{array}{cc} 1 & 0\\ 
\!\!\!-B &  1\end{array}\right)
= 
\left[\exp(R)\right]^M{}_S\cdot\mathcal{Q}^S{}_P\cdot\left[\exp(-R)\right]^P{}_N
~,
}
where we took  into account that $\exp(R)=1+R$, as follows from (\ref{r}), and we have noted that 
\eq{\mathcal{Q}^M{}_N=\mathcal{G}^{MS}\cdot\mathcal{Q}_S{}^P\cdot\mathcal{G}_{PN}=
\left(\begin{array}{cc} 0 & ~~\!g^{-1}\\ 
g & 0 \end{array}\right)~.
}
On the other hand it follows from the discussion below (\ref{40}) that $R$ is an element of $\mathfrak{so}(D,D)$, in particular a $B$-transform. In other words (\ref{86}) is 
saying that the most general $\mathcal{Q}'$ is obtained from the $\mathcal{Q}$ coming from supergravity by an $SO(D,D)$ $B$-transform. Correspondingly the most 
general GACS are obtained from those coming from supergravity, cf.~(\ref{ee}), by the same $SO(D,D)$ $B$-transform: this is precisely the content of eqn.~(\ref{r1}). Finally recall that the spinorial action of the $\mathfrak{so}(D,D)$ $B$-transform is given by,
\eq{\frac{1}{2}R_{MN}\Gamma^{MN}=\frac{1}{2}B_{ij}\d x^{i}\wedge 
\d x^{j}\wedge=B\wedge~,}
cf.~below (\ref{244}), which indeed exponentiates to (\ref{r2}).

Let us make one further comment about (\ref{r2}). In general the $B$-field 
appearing on the right-hand side of that equation need not be globally 
defined on the  manifold $\mcal$. Let $U_i$, $U_j$ be overlapping open patches of $\mcal$, and let $B_i$, $B_j$ be the value 
of $B$ on $U_i$, $U_j$ respectively. On $U_i\cap U_j$ we have,
\eq{\label{g1}B_i=B_j+\d\Lambda_{(ij)}~,}
where $\Lambda_{(ij)}\in U_i\cap U_j$ is a one-form defined on the double overlap. Indeed 
in supergravity we identify the $B$ field with the `potential' of the globally-defined Neveu-Schwarz threeform $H=\d B$, and 
the above patching leaves $H$ invariant.  
 On triple overlaps  
$U_i\cap U_j\cap U_k$ the consistency condition $0=(B_i-B_j)+(B_j-B_k)+(B_k-B_i)$  then leads to:
\eq{\d\left(\Lambda_{(ij)}+\Lambda_{(jk)}+\Lambda_{(ki)}\right)=0~.}
Taking into account that $U_i\cap U_j\cap U_k$ is topologically trivial, the above is equivalent to,
\eq{\label{g2}\Lambda_{(ij)}+\Lambda_{(jk)}+\Lambda_{(ki)}=\d\Lambda_{(ijk)}~,}
where $\Lambda_{(ijk)}$ is a function on the triple overlap. Eqs.~(\ref{g1}),(\ref{g2}) mean that $B$ is what is known as a {\it connection on a gerbe}, a higher-dimensional generalization of a connection one-form. 

As a consequence of the nontrivial patching (\ref{g1}), the pure spinors 
$e^B\wedge\Psi_a$ are not in general globally defined on the generalized tangent bundle $T\oplus T^*$ over $\mcal$. Indeed since the supergravity parameters $\eta_a$, $a=1,2$, are globally 
defined pure spinors on $\mcal$ (i.e. global sections of the spin bundle over $\mcal$), the generalized pure spinors $\Psi_a$, cf.~(\ref{ff}), are  
global sections of the generalized tangent bundle $T\oplus T^*$ with 
transition functions in $SO(D)$. On the double overlap we thus have,
\eq{\left.\big(e^B\wedge\Psi_a\big)\right|_{U_i}=e^{\d\Lambda_{(ij)}}\wedge
\left.\big(e^B\wedge\Psi_a\big)\right|_{U_j}~.}
However, we may define a {\it twisted generalized bundle} $E$ over $\mcal$ whose fibers are $T\oplus T^*$ as before, 
but whose transition functions across patches include a $B$-transform in $SO(D,D)$,
\eq{\left.\Psi_a\right|_{U_i}=e^{-\d\Lambda_{(ij)}}\wedge\left.\Psi_a\right|_{U_j}~,}
precisely of the form to counterbalance the effect of the nontrivial patching of the $B$ field. On the twisted bundle $E$ the pure spinor $e^B\wedge\Psi_a$ is thus globally defined. The fact that the transition functions of $E$ `know about' the $B$ field is sometimes referred to as `geometrizing the $B$ field'. Extending the same procedure to the remaining supergravity fields leads to the {\it exceptional generalized geometry} \cite{Pacheco:2008ps} which is beyond the scope of these notes.

\color{black}

The existence of a pair of commuting, metric-compatible almost complex structures allows us to decompose the space of polyforms/generalized spinors 
into $+i(k,l)$ eigenspaces, $U_{k,l}$, of $(\mathcal{I}_1,\mathcal{I}_2)$. Explicitly this means that the spinorial action of $\mathcal{I}$ is 
given by,
\eq{\frac12 \mathcal{I}_1^{MN}\Gamma_{MN}u_{k,l}= ik~\!u_{k,l}~;~~~ 
\frac12 \mathcal{I}_2^{MN}\Gamma_{MN}u_{k,l}= il~\!u_{k,l}~,}
for $u_{k,l}$ a basis of $U_{k,l}$.

\color{mygreen}

This is completely analogous to the ordinary case: As we have seen, spinors 
are built from the Fock vacuum (the pure spinor) $\eta$ by applying a number of 
creation operators. The space of spinors can be therefore decomposed according 
to the `occupation level' of each state, i.e. the number of creation operators one needs to apply to the vacuum in order to obtain each state. Explicitly we have,
\eq{\label{sr}\gamma^-\eta=0~;~~~u_{2p-{n}}:=\gamma_{(p)-}
\eta^c\sim\gamma_{({n}-p)+}\eta~;~~~
\frac12 {J}_{ij}\gamma^{ij}u_{k}= ik~\!u_{k}~,}
where $p$ runs from zero to $n=D/2$, while $k$ runs from $-n$ to $+n$ in steps of two. I.e., $p=0,1,\dots, n$~;~~ $k=2p-{n}=-n,-n+2,\dots, n$. In the equation  above  $\gamma_{(p)\pm}$ stands for a product of $p$ gamma matrices holomorphic, antiholomorphic with respect to $\eta$ respectively.

\color{mygreen}

The second relation in (\ref{sr}) can be thought of as a {\it holomorphic Hodge duality} between 
antisymmetric products of gamma matrices. Indeed let us consider the case $D=10$ for concreteness. Using the Fierzing techniques that were explained earlier, the gamma matrices can be shown to obey the following relations \cite{Prins:2014ona},
\eq{\spl{\label{b11}
\g_m\eta      &= (\Pi^+)_{m}{}^{n}\g_n\eta\\
\g_{mn}\eta   &= iJ_{mn}\eta + (\Pi^+)_{[m}^{\phantom{[m}p}(\Pi^+)_{n]}^{\phantom{[m}q} \g_{pq}   \eta  \\
\g_{mnp}\eta  &= 3iJ_{[mn}\g_{p]}\eta  +\frac{1}{8}\Omega_{mnpqr}\g^{qr}\eta^c\\
\g_{mnpq}\eta &= -3J_{[mn}J_{pq]}\eta + 6i J_{[mn}(\Pi^+)_{p}^{\phantom{[m}r}(\Pi^+)_{q]}^{\phantom{[m}s} \g_{rs}   \eta
                 - \frac{1}{2} \O_{mnpqr} \g^r \eta^c \\
\g_{mnpqr}\eta &=  - \O_{mnpqr} \eta^c + \frac{5i}{4} J_{[mn}\O_{pqrst}\g^{st} \eta^c - 15 J_{[mn}J_{pq}\g_{r]} \eta
~.
}}
Projecting both sides with $\Pi^+$ onto the holomorphic part, the first two equations of (\ref{b11}) become identically satisfied. From the remaining equations we obtain,
\eq{\spl{\label{b12}
\g_{mnp+}\eta  &= +\frac{1}{2^22!}\Omega_{mnp}{}^{qr}\g_{qr-}\eta^c\\
\g_{mnpq+}\eta &= - \frac{1}{2} \O_{mnpq}{}^{r} \g_{r-} \eta^c \\
\g_{mnpqr+}\eta &=  - \O_{mnpqr} \eta^c 
~.
}}
The above relations are indeed of the form of a holomorphic Hodge duality, with $\Omega$ playing the role of a `holomorphic $\varepsilon$-tensor'. 
Moreover, the equations above can be inverted by contracting with $\O^*_{ij}{}^{mnp}$, $\O^*_{i}{}^{mnpq}$ and $\O^{*mnpqr}$ respectively, taking the complex conjugate, and using the relations
\eq{\spl{ \label{eq:omegas}
\frac{1}{2^55!}~&\Omega_{vwxyz}\Omega^{*vwxyz}=1\\
\frac{1}{2^5 4!}~&\Omega_{awxyz}\Omega^{*mwxyz}
=(\Pi^+)_{a}{}^{m}\\
\frac{1}{ 2^5 12}~&\Omega_{abxyz}\Omega^{*mnxyz}
=(\Pi^+)_{[a}{}^{m}(\Pi^+)_{b]}{}^{n}~,
}}
which are holomorphic analogues of the usual $\varepsilon^{m\cdots}\varepsilon_{n\cdots}\sim\delta^{m\cdots}_{n\cdots}$ 
relations. We thus obtain,
\eq{\spl{\label{b13}
\g_{mn+}\eta  &= +\frac{1}{2^33!}\Omega_{mn}{}^{pqr}\g_{pqr-}\eta^c\\
\g_{m+}\eta  &= -\frac{1}{2^44!}\Omega_{m}{}^{npqr}\g_{npqr-}\eta^c\\
\eta  &= -\frac{1}{2^55!}\Omega^{mnpqr}\g_{mnpqr-}\eta^c
~.
}}
Finally, to show the third equation in (\ref{sr}) we first calculate,
\eq{\spl{
[\underline{J},\g^{\pm}_q ]&=
\frac12 J_{mn}\underbrace{\big(\Pi^{\pm}\big)_q{}^r}_{
\frac12(\delta^r_q\mp iJ_q{}^r)
} \underbrace{[\g^{mn},\g_r ]}_{
4\g^{[m}\delta^{n]}_r
}\\
&=\mp i\underbrace{(g_{qm}\mp iJ_{qm})}_{2\big(\Pi^{\pm}\big)_{qm}}\g^m \\
&=\mp 2i\gamma^{\pm}_q
~.
}}
Using the above we obtain,
\eq{\spl{
\underline{J}~\!u_{{n}-2k}&=
\underline{J}\g^+_{m_1}\cdots\g^+_{m_k}\eta\\
&=\left(
[\underline{J},\g^{+}_{m_1} ]+\g^{+}_{m_1}\underline{J}
\right)\g^+_{m_2}\cdots\g^+_{m_k}u_{{n}-2k}\\
&=-2i~\!u_{{n}-2k}+\g^{+}_{m_1}\underline{J}\g^+_{m_2}\cdots
\g^+_{m_k}\eta\\
&\vdots\\
&=-2ik~\! u_{{n}-2k}+\g^{+}_{m_1}\cdots
\g^+_{m_k}\underbrace{\underline{J}\eta}_{in~\!\!\eta}\\
&=i ({n}-2k)~\!u_{{n}-2k}
~.}}
In the penultimate line above we used the second line of (\ref{b11}), which can be seen to hold in any spacetime dimension $D$, so that:
\eq{
\underline{J}\eta=\frac12 J^{ij}\left(i J_{ij}+\g_{ij+} \right)\eta=in~\!\eta
~,}
since $J_{ml}J^{lp}=-\delta_m^p$, and $J_{ij}$ is a (1,1)-tensor with respect to 
the almost complex structure whereas $\g_{ij+}$ is (2,0) so its contraction 
with $J$ vanishes.

In the generalized case the subspaces $u_k$, the pure spinor $\eta$ and the annihilation operators 
$\gamma^-$ are replaced by $u_{k,l}$, $\Psi_{1,2}$ and $L_{1,2}^{(\pm)}$ respectively:
\eq{u_k\rightarrow u_{k,l}~;~~~\eta\rightarrow\Psi_1~;~~~\gamma^-\rightarrow L_1^{(\pm)}~,}
equivalently,
\eq{u_l\rightarrow u_{k,l}~;~~~\eta\rightarrow\Psi_2~;~~~\gamma^-\rightarrow L_2^{(\pm)}~.}

\color{black}

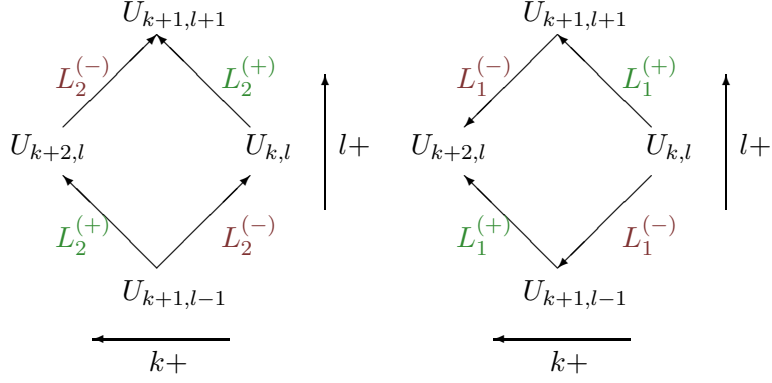
\begin{figure}\color{black}
\begin{picture}(400,90)(-62,40)
\put(202,185){$U_{k+1,l+1}$}

\put(177,160){$\color{myred}L^{(-)}_1$}

\put(239,160){$\color{mygreen}L^{(+)}_1$}

\put(215,180){\vector(-1,-1){35}}\put(250,145){\vector(-1,1){35}}

\put(160,135){$U_{k+2,l}$}\put(248,135){$U_{k,l}$}

\put(215,92){\vector(-1,1){35}}\put(250,127){\vector(-1,-1){35}}

\put(177,100){$\color{mygreen}L^{(+)}_1$}

\put(239,100){$\color{myred}L^{(-)}_1$}

\put(278,114){\vector(0,1){51}}

\put(283,135){$l+$}

\put(242,65){\vector(-1,0){51}}

\put(212,53){$k+$}

\put(202,80){$U_{k+1,l-1}$}


\put(52,185){$U_{k+1,l+1}$}

\put(27,160){$\color{myred}L^{(-)}_2$}

\put(89,160){$\color{mygreen}L^{(+)}_2$}

\put(30,145){\vector(1,1){35}}\put(100,145){\vector(-1,1){35}}

\put(10,135){$U_{k+2,l}$}\put(98,135){$U_{k,l}$}

\put(65,92){\vector(-1,1){35}}\put(65,92){\vector(1,1){35}}

\put(27,100){$\color{mygreen}L^{(+)}_2$}

\put(89,100){$\color{myred}L^{(-)}_2$}

\put(128,114){\vector(0,1){51}}

\put(133,135){$l+$}

\put(92,65){\vector(-1,0){51}}

\put(62,53){$k+$}

\put(52,80){$U_{k+1,l-1}$}

\end{picture}
\caption{\label{f1}The action of $L_{1,2}$ on the generalized subspaces $U_{k,l}$:  
$L_1^{(\pm)}$, $\overline{L_1^{(\pm)}}$ increases, resp.~decreases the value of $k$ by one. Similarly, $L_2^{(\pm)}$, $\overline{L_2^{(\pm)}}$ increases, resp.~decreases the value of $l$ by one. Note also the identifications 
$L_1^{(+)}= L_2^{(+)}$ and $L_1^{(-)}= \overline{L_2^{(-)}}$, cf.~(\ref{25}).}
\end{figure}

The $U_{k,l}$ subspaces can be obtained by applying a number of creation operators $\overline{L^{\pm}_1}$ on $\Psi_1$,  or $\overline{L^{\pm}_2}$ on $\Psi_2$:
\eq{\spl{\label{e1}
U_{k,l}&\sim\left(\overline{L^+_1}\right)^{\frac{{n}-(k+l)}{2}}
\left(\overline{L^-_1}\right)^{\frac{{n}-(k-l)}{2}}\Psi_1\\
&\sim
\gamma_{\frac{{n}-(k+l)}{2}+}\Psi_1
~\gamma_{\frac{{n}-(k-l)}{2}-}
~,}}
equivalently,
\eq{\spl{\label{e2}
U_{k,l}&\sim\left(\overline{L^+_2}\right)^{\frac{{n}-(k+l)}{2}}
\left(\overline{L^-_2}\right)^{\frac{{n}+(k-l)}{2}}\Psi_2\\
&\sim
\gamma_{\frac{{n}-(k+l)}{2}+}\Psi_2
~\gamma_{\frac{{n}+(k-l)}{2}+}
~,}}
where  a $\gamma_{(p)\pm}$ on the right of $\Psi_1$ stands for a product of $p$ gamma matrices holomorphic or antiholomorphic with respect to $\eta_2$, while a $\gamma_{(p)\pm}$ on the left of $\Psi_1$ stands for a product of $p$ gamma matrices holomorphic or antiholomorphic with respect to $\eta_1$. 
Figure \ref{f1} shows how the generalized creation, annihilation operators move us from 
one subspace to another.

\color{mygreen}

The equivalence between (\ref{e1}), (\ref{e2}) is established by taking 
into account the holomorphic Hodge-duality property of the gamma matrices,
\eq{\gamma_{(p)-}\eta^c\sim  \gamma_{({n}-p)+}\eta~,}
which implies in particular,
\eq{
\underbrace{\eta_1\otimes\widetilde{\eta_2^c}}_{\Psi_1}~\gamma_{\frac{{n}-(k-l)}{2}-}\sim
\underbrace{\eta_1\otimes\widetilde{\eta_2}}_{\Psi_2}~\gamma_{\frac{{n}+(k-l)}{2}+}
~.}
The pure spinors $\Psi$, $\Psi^*$ themselves can serve as bases for the 
subspaces corresponding to the generalized Fock vaccum and its adjoint (with respect to the Mukai pairing):
\eq{\spl{
U_{{n},0}&\sim\Psi_1~;~~~U_{-{n},0}
\sim \gamma_{(n)+}\Psi_1~\gamma_{(n)-}
\sim\Psi^*_1
\\
U_{0,{n}}&\sim\Psi_2~;~~~U_{0,-{n}}
\sim \gamma_{(n)+}\Psi_2~\gamma_{(n)+}
\sim\Psi^*_2
~.}}

\color{black}

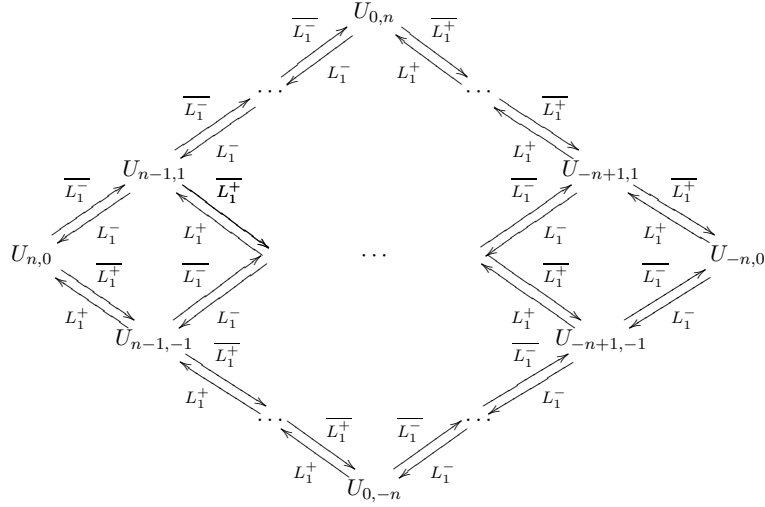
\begin{figure}\color{black}
\begin{equation*}\label{genhodgediamond2}
\resizebox{.8\hsize}{!}{
\centerline{
\xymatrix{
&&&  U_{0,{n}}\ar@{->}[ld]<0.5ex>^{L_1^-}\ar@{->}[rd]<0.5ex>^{\overline{L_1^+}} &&& \\
&&\cdots\ar@{->}[ru]<0.5ex>^{\overline{L_1^-}}\ar@{->}[ld]<0.5ex>^{L_1^-} &&\cdots\ar@{->}[rd]<0.5ex>^{\overline{L_1^+}}\ar@{->}[lu]<0.5ex>^{L_1^+}&&\\
&U_{{n}-1,1}\ar@{->}[rd]<0.5ex>^{\overline{L_1^+}}\ar@{->}[ld]<0.5ex>^{L_1^-}\ar@{->}[ru]<0.5ex>^{\overline{L_1^-}}\ar@{->}[rd]<0.5ex>^{\overline{L_1^+}}&& &&U_{-{n}+1,1}\ar@{->}[lu]<0.5ex>^{L_1^+}\ar@{->}[ld]<0.5ex>^{L_1^-}\ar@{->}[rd]<0.5ex>^{\overline{L_1^+}}&\\
U_{{n},0}\ar@{->}[rd]<0.5ex>^{\overline{L_1^+}}\ar@{->}[ru]<0.5ex>^{\overline{L_1^-}}&&\ar@{->}[lu]<0.5ex>^{L_1^+}\ar@{->}[ld]<0.5ex>^{L_1^-}   &\cdots &\ar@{->}[rd]<0.5ex>^{\overline{L_1^+}}\ar@{->}[ru]<0.5ex>^{\overline{L_1^-}}&&U_{-{n},0}\ar@{->}[lu]<0.5ex>^{L_1^+}\ar@{->}[ld]<0.5ex>^{L_1^-}\\
&U_{{n}-1,-1}\ar@{->}[lu]<0.5ex>^{L_1^+}\ar@{->}[rd]<0.5ex>^{\overline{L_1^+}}\ar@{->}[ru]<0.5ex>^{\overline{L_1^-}}&&
&&U_{-{n}+1,-1}\ar@{->}[ru]<0.5ex>^{\overline{L_1^-}}\ar@{->}[lu]<0.5ex>^{L_1^+}\ar@{->}[ld]<0.5ex>^{L_1^-}&\\
&&\ldots\ar@{->}[lu]<0.5ex>^{L_1^+}\ar@{->}[rd]<0.5ex>^{\overline{L_1^+}}&&\ldots\ar@{->}[ru]<0.5ex>^{\overline{L_1^-}}\ar@{->}[ld]<0.5ex>^{L_1^-}&&\\
&&&  U_{0,-{n}}\ar@{->}[ru]<0.5ex>^{\overline{L_1^-}}\ar@{->}[lu]<0.5ex>^{L_1^+}  &&&
}}
}
\end{equation*}
\caption{\label{f3}The generalized Hodge diamond. The arrows indicate the 
action of the $\overline{L_1}$ , $L_1$ creation, annihilation operators 
respectively. The generalized vacuuum, represented by $U_{{n},0}\sim\Psi_1$, is annihilated 
by $L_1^{\pm}$. All generalized subspaces $U_{k,l}$ can be reached by 
a finite number of successive actions of the creation operators $\overline{L_1^{\pm}}$ on the vacuum $U_{{n},0}$.}
\end{figure}

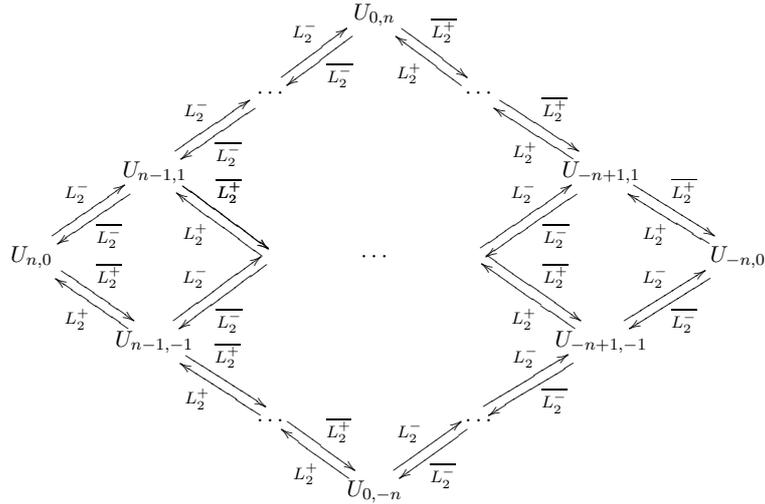
\begin{figure}\color{black}
\begin{equation*}\label{genhodgediamond1}
\resizebox{.8\hsize}{!}{
\centerline{
\xymatrix{
&&&  U_{0,{n}}\ar@{->}[ld]<0.5ex>^{\overline{L_2^-}}\ar@{->}[rd]<0.5ex>^{\overline{L_2^+}} &&& \\
&&\cdots\ar@{->}[ru]<0.5ex>^{L_2^-}\ar@{->}[ld]<0.5ex>^{\overline{L_2^-}} &&\cdots\ar@{->}[rd]<0.5ex>^{\overline{L_2^+}}\ar@{->}[lu]<0.5ex>^{L_2^+}&&\\
&U_{{n}-1,1}\ar@{->}[rd]<0.5ex>^{\overline{L_2^+}}\ar@{->}[ld]<0.5ex>^{\overline{L_2^-}}\ar@{->}[ru]<0.5ex>^{L_2^-}\ar@{->}[rd]<0.5ex>^{\overline{L_2^+}}&& &&U_{-{n}+1,1}\ar@{->}[lu]<0.5ex>^{L_2^+}\ar@{->}[ld]<0.5ex>^{\overline{L_2^-}}\ar@{->}[rd]<0.5ex>^{\overline{L_2^+}}&\\
U_{{n},0}\ar@{->}[rd]<0.5ex>^{\overline{L_2^+}}\ar@{->}[ru]<0.5ex>^{L_2^-}&&\ar@{->}[lu]<0.5ex>^{L_2^+}\ar@{->}[ld]<0.5ex>^{\overline{L_2^-}}   &\cdots &\ar@{->}[rd]<0.5ex>^{\overline{L_2^+}}\ar@{->}[ru]<0.5ex>^{L_2^-}&&U_{-{n},0}\ar@{->}[lu]<0.5ex>^{L_2^+}\ar@{->}[ld]<0.5ex>^{\overline{L_2^-}}\\
&U_{{n}-1,-1}\ar@{->}[lu]<0.5ex>^{L_2^+}\ar@{->}[rd]<0.5ex>^{\overline{L_2^+}}\ar@{->}[ru]<0.5ex>^{L_2^-}&&
&&U_{-{n}+1,-1}\ar@{->}[ru]<0.5ex>^{L_2^-}\ar@{->}[lu]<0.5ex>^{L_2^+}\ar@{->}[ld]<0.5ex>^{\overline{L_2^-}}&\\
&&\ldots\ar@{->}[lu]<0.5ex>^{L_2^+}\ar@{->}[rd]<0.5ex>^{\overline{L_2^+}}&&\ldots\ar@{->}[ru]<0.5ex>^{L_2^-}\ar@{->}[ld]<0.5ex>^{\overline{L_2^-}}&&\\
&&&  U_{0,-{n}}\ar@{->}[ru]<0.5ex>^{L_2^-}\ar@{->}[lu]<0.5ex>^{L_2^+}  &&&
}}
}
\end{equation*}
\caption{\label{f4}The generalized Hodge diamond. The arrows indicate the 
action of the $\overline{L_2}$ , $L_2$ creation, annihilation operators 
respectively. The generalized vacuuum, represented by $U_{0,{n}}\sim\Psi_2$, is annihilated 
by $L_2^{\pm}$. All generalized subspaces $U_{k,l}$ can be reached by 
a finite number of successive actions of the creation operators $\overline{L_2^{\pm}}$ on the vacuum $U_{0,{n}}$.}
\end{figure}

Figure \ref{f3} shows the generalized Hodge diamond, where we have indicated how the $L_1$ creation, annihilation operators move us between subspaces. In figure \ref{f4} we show the same Hodge diamond indicating instead the action of the  $L_2$ creation, annihilation operators.

\begin{figure}
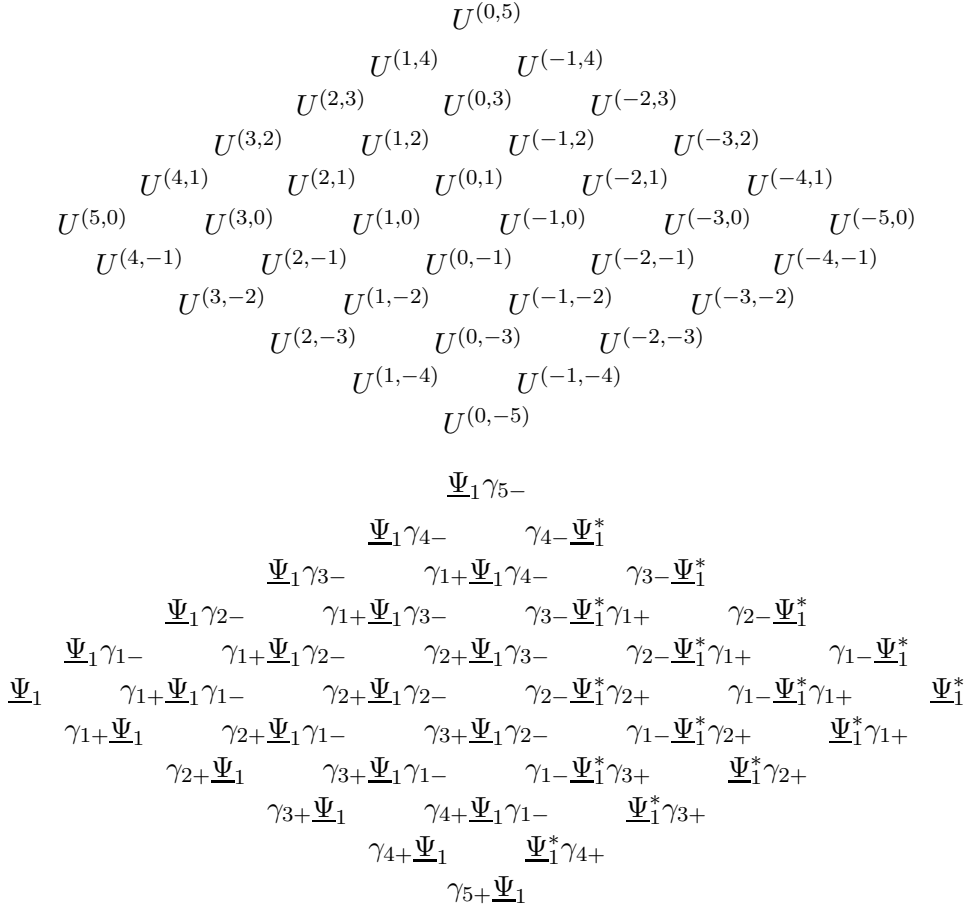
\color{black}
{
\eq{\label{hodge1}
\begin{array}{c}\vspace{.1cm}
U^{(0,5)}\\
U^{(1,4)}\hspace{1cm}
U^{(-1,4)}\\
U^{(2,3)}\hspace{1cm}U^{(0,3)}\hspace{1cm}
U^{(-2,3)}\\
U^{(3,2)}\hspace{1cm}U^{(1,2)}\hspace{1cm}
U^{(-1,2)}\hspace{1cm}U^{(-3,2)}\\
U^{(4,1)}\hspace{1cm}U^{(2,1)}\hspace{1cm}
U^{(0,1)}\hspace{1cm}
U^{(-2,1)}\hspace{1cm}U^{(-4,1)}\\
U^{(5,0)}\hspace{1cm}U^{(3,0)}\hspace{1cm}
U^{(1,0)}\hspace{1cm}
U^{(-1,0)}\hspace{1cm}U^{(-3,0)}
\hspace{1cm}U^{(-5,0)}\\
U^{(4,-1)}\hspace{1cm}U^{(2,-1)}\hspace{1cm}
U^{(0,-1)}\hspace{1cm}
U^{(-2,-1)}\hspace{1cm}U^{(-4,-1)}\\
U^{(3,-2)}\hspace{1cm}U^{(1,-2)}\hspace{1cm}
U^{(-1,-2)}\hspace{1cm}U^{(-3,-2)}\\
U^{(2,-3)}\hspace{1cm}U^{(0,-3)}\hspace{1cm}
U^{(-2,-3)}\\
U^{(1,-4)}\hspace{1cm}
U^{(-1,-4)}\\
U^{(0,-5)}
\end{array}
\nn}
\eq{\label{hodge2}
\begin{array}{c}\vspace{.1cm}
\underline{\Psi}_1\gamma_{5-}\\
\underline{\Psi}_1\gamma_{4-}\hspace{1cm}
\gamma_{4-}\underline{\Psi}_1^*\\
\underline{\Psi}_1\gamma_{3-}\hspace{1cm}\gamma_{1+}\underline{\Psi}_1\gamma_{4-}\hspace{1cm}
\gamma_{3-}\underline{\Psi}_1^*\\
\underline{\Psi}_1\gamma_{2-}\hspace{1cm}\gamma_{1+}\underline{\Psi}_1\gamma_{3-}\hspace{1cm}
\gamma_{3-}\underline{\Psi}_1^*\gamma_{1+}\hspace{1cm}\gamma_{2-}\underline{\Psi}_1^*\\
\underline{\Psi}_1\gamma_{1-}\hspace{1cm}\gamma_{1+}\underline{\Psi}_1\gamma_{2-}\hspace{1cm}
\gamma_{2+}\underline{\Psi}_1\gamma_{3-}\hspace{1cm}
\gamma_{2-}\underline{\Psi}_1^*\gamma_{1+}\hspace{1cm}\gamma_{1-}\underline{\Psi}_1^*\\
\underline{\Psi}_1\hspace{1cm}\gamma_{1+}\underline{\Psi}_1\gamma_{1-}\hspace{1cm}
\gamma_{2+}\underline{\Psi}_1\gamma_{2-}\hspace{1cm}
\gamma_{2-}\underline{\Psi}_1^*\gamma_{2+}\hspace{1cm}\gamma_{1-}\underline{\Psi}_1^*\gamma_{1+}
\hspace{1cm}\underline{\Psi}_1^*\\
\gamma_{1+}\underline{\Psi}_1\hspace{1cm}\gamma_{2+}\underline{\Psi}_1\gamma_{1-}\hspace{1cm}
\gamma_{3+}\underline{\Psi}_1\gamma_{2-}\hspace{1cm}
\gamma_{1-}\underline{\Psi}_1^*\gamma_{2+}\hspace{1cm}\underline{\Psi}_1^*\gamma_{1+}\\
\gamma_{2+}\underline{\Psi}_1\hspace{1cm}\gamma_{3+}\underline{\Psi}_1\gamma_{1-}\hspace{1cm}
\gamma_{1-}\underline{\Psi}_1^*\gamma_{3+}\hspace{1cm}\underline{\Psi}_1^*\gamma_{2+}\\
\gamma_{3+}\underline{\Psi}_1\hspace{1cm}\gamma_{4+}\underline{\Psi}_1\gamma_{1-}\hspace{1cm}
\underline{\Psi}_1^*\gamma_{3+}\\
\gamma_{4+}\underline{\Psi}_1\hspace{1cm}
\underline{\Psi}_1^*\gamma_{4+}\\
\gamma_{5+}\underline{\Psi}_1
\end{array}\nn
}
}
\caption{\label{f5}The generalized Hodge diamond in the maximal dimension $D=10$. In the second diagram we have indicated explicit bases for the $U_{k,l}$ subspaces in terms of generalized spinors built from $\Psi_1$, cf.~(\ref{e1}). A similar diamond could be built using $\Psi_2$, cf.~(\ref{e2}). A $\gamma_{p\pm}$ on the right of $\Psi_1$ stands for a product of $p$ gamma matrices holomorphic or antiholomorphic with respect to $\eta_2$, while a $\gamma_{p\pm}$ on the left of $\Psi_1$ stands for a product of $p$ gamma matrices holomorphic or antiholomorphic with respect to $\eta_1$. 
}
\end{figure}

In figure \ref{f5} we show the generalized Hodge diamond in the maximal dimension $D=10$. We have also used (\ref{e1}) to explicitly represent the subspaces $U_{k,l}$. Of course we could equally well have used (\ref{e2}) 
instead.

Note in particular that the subspaces are orthogonal with respect to the Mukai pairing:
\eq{\label{ortho}\langle u_{k,l},u_{p,q}\rangle\propto\delta_{k+p}\delta_{l+q}~,}
where $u_{k,l}$ is a basis of $U_{k,l}$.
The upshot is that any polyform can be decomposed as follows,
\eq{\label{polydec}\Phi=\sum_{k=-{n}}^{{n}}\left(\Phi_{k,|k|-{n}}+
\Phi_{k,|k|-{n}+2}+\dots+\Phi_{k,{n}-|k|}\right)
~,}
where $\Phi_{k,l}\in U_{k,l}$ is the projection of $\Phi$ onto the 
subspace $U_{k,l}$,
\eq{\label{proj}\Phi_{k,l}\propto\langle u_{-k,-l},\Phi\rangle~\! u_{k,l} ~.}
This has the important practical consequence that in order to prove the validity of a polyform equation, i.e. an equation of the form $\Phi=0$ with $\Phi$ a polyform, it is sufficent (and necessary) to show that $\Phi_{k,l}=0$ for all $k$, $l$. We will come back to this point later on.

\color{mygreen}

Eqn.~(\ref{ortho}) can easily be shown by using the bispinor expresion (\ref{7}) for 
the Mukai pairing. For example in the representation (\ref{e1}) we find, 
substituting the expression of $\Psi_1$ from (\ref{ff}), 
\eq{\spl{
\langle u_{k,l},u_{p,q}\rangle/
\mathrm{vol}_{D}
&\propto\mathrm{tr}\left(
\underline{\widetilde{u_{k,l}}}
\gamma_{D+1}\underline{u_{p,q}}
\right)\\
&\propto\mathrm{tr}\left(\gamma_{\frac{{n}-(k-l)}{2}-}
\eta^c_2\otimes\widetilde{\eta}_1~
\gamma_{\frac{{n}-(k+l)}{2}+}
~
\gamma_{\frac{{n}-(p+q)}{2}+}\eta_1\otimes\widetilde{\eta^c_2}
~\gamma_{\frac{{n}-(p-q)}{2}-}
\right)\\
&\propto
\left(
\widetilde{\eta^c_2}~\gamma_{\frac{D-(k+p)+(l+q)}{2}-}\eta^c_2\right)
\left(
\widetilde{\eta}_1~\gamma_{\frac{D-(k+p)-(l+q)}{2}+}\eta_1\right)\\
&\propto\delta_{k+p}\delta_{l+q}
~,}}
where to go from the penultimate to the last line we took into account that $\eta_1$, $\eta_2$ are pure spinors hence all their bilinears vanish except for the one of order $D/2$, cf. (\ref{ch}).

\color{black}

We need one last element in order to complete this review of generalized geometry. Just as in the ordinary case, there is a notion of integrability of 
GACS. This is based on the so-called ($H$-twisted) Courant bracket, the generalized analogue of the Lie bracket. It can be shown that the integrability of a GACS $\mathcal{I}$ can be expressed equivalently as a differential condition on the associated generalized pure spinor $\Psi$:
\eq{\label{int}\d_H\Psi = {V}^M\Gamma_M\Psi~,}
for some $V\in T\oplus T^*$. In the equation above $\d_H$ is the $H$-twisted differential 
$\d_H:=\d+H\wedge$; it is nilpotent for $H$ closed. A manifold $\mcal_{2n}$ with an integrable almost complex structure 
is called {\it generalized complex}; it is locally equivalent to $\mathbb{C}^q\times(\mathbb{R}^{2(n-q)},J)$ up to a $B$-transform, with $J$ the standard 
symplectic structure. Thus generalized complex geometry can be said to be an interpolation between complex and symplectic geometries. The integer $q$ is called the {\it type}, and need not be constant over $\mcal_{2n}$.

A generalized complex manifold with a pure spinor such that the right hand-side of (\ref{int}) vanishes is called {\it generalized Calabi-Yau} (GCY), according to the definition of Hitchin in \cite{hitch}.\footnote{The reader should be aware that Hitchin's definition of GCY differs from that of Gualtieri's in \cite{gual}.} As we will see, supersymmetry imposes that all pure backgrounds of type II super-gravity that we consider here must have internal spaces which are GCY. 

If the manifold is generalized complex, it can be shown  that the $H$-twisted 
differential sends $k$-polyforms to $(k\pm 1)$-polyforms:
\eq{\label{fnl}
\d_H \Phi_{k} =\left(\d_H\Phi\right)_{k+1}
+\left(\d_H\Phi\right)_{k-1}
~.}
We can thus introduce {\it generalized Dolbeault operators}, 
${\partial}_H^{\mathcal{I}}$, $\bar{\partial}_H^{\mathcal{I}}$, associated with the integrable GACS $\mathcal{I}$:
\eq{\label{dolbeault}\partial_H^{\mathcal{I}}\Phi_k:=\left(\d_H\Phi\right)_{k+1}~,
~~~~~\bar{\partial}_H^{\mathcal{I}}\Phi_k:=\left(\d_H\Phi\right)_{k-1}~,}
and a {\it generalized $\d^c$ \!operator}, 
\eq{\label{gd}\d_H^{\mathcal{I}}:= i\big(\bar{\partial}_H^{\mathcal{I}}-\partial_H^{\mathcal{I}}\big)
~.}

\color{mygreen}

Suppose that $\Phi_k\in U_k$, where $U_k$ denotes the 
$+ik$-eigenspace of the integrable GACS $\mathcal{I}$. For concreteness   
let us take $\mathcal{I}=\mathcal{I}_1$, so that $U_k=\sum_l\oplus U_{k,l}$. 
To show (\ref{fnl}) first note that we can use (\ref{e1}), (\ref{proj}) to expand $\Phi_k$ as follows,
\eq{\underline{\Phi_k}=\sum_{p+q={n}-k}
C^{m_1\cdots m_pr_1\cdots r_q}\g^+_{m_1\cdots m_p}\underline{\Psi_1}
\g^-_{r_1\cdots r_q}
~,}
for some coefficients $C$. On the other hand, 
using (\ref{ga}),(\ref{p}),
\eq{\spl{\label{here}
\d_H\Phi_k &= {V}^M\Gamma_M\Phi_k\\
&=(\iota_a+b\wedge)\Phi_k\\
&=
\frac12 a^m\big(\gamma_m\underline{\Phi_k}-(-1)^{|\Phi|}\underline{\Phi_k}\gamma_m\big)
+
\frac12 b_m
\big(\gamma^m\underline{\Psi}+(-1)^{|\Phi|}\underline{\Phi_k}\gamma^m\big)
~.
}}
Moreover decomposing $\g_m=\g^+_m+\g^-_m$ and taking into account that
\eq{
\g^{\pm}_m\underline{\Phi_k}\subset U_{k\mp 1}~;~~~
\underline{\Phi_k}\g^{\pm}_m\subset U_{k\pm 1}
~,}
as can be seen from (\ref{e1}), 
we arrive at (\ref{fnl}).

\color{black}

\section{Part II: pure backgrounds in various dimensions}

In this part we report on some past and recent results on pure backgrounds of type II supergravity in various dimensions, and the supersymmetry/calibrations correspondence.

\subsection{Calibrations and supersymmetry}

An ordinary {\it calibration} \cite{Harvey:1982xk} of degree $l$ on a Riemannian manifold $\mcal$ is given by a closed $l$-form $\omega$ on $\mcal$, $\d\omega=0$, such that at each point $p\in\mcal$ 
and for each oriented, $l$-dimensional subspace $T\subset T_p\mathcal{M}$, the following inequality holds,
\eq{\omega(T)\leq \sqrt{g|_T}~.}
The pullback onto $T$ of the metric $g$ of $\mcal$ and the evaluation, $\omega(T)$, of $\omega$ on $T$ are defined as follows: Suppose $\{t_a,~a=1,\dots l\}$, is an oriented basis of $T$. Then $(g|_T)_{ab}:=g_{mn}t_a^mt^n_b$ and $\omega(T):=\iota_{t_1}\cdots\iota_{t_l}\omega|_p$. Moreover we demand that 
at each point $p\in\mcal$, the inequality is saturated for some oriented subspace of $T\subset T_p\mathcal{M}$,
\eq{\{ T\subset T_p\mathcal{M}~|~  \omega(T)= \sqrt{g|_T}  \}\neq \emptyset~.}
Suppose now that there is an oriented submanifold $\Sigma\subset\mathcal{M}$, parameterized by coordinates $\{\sigma^a,~a=1,\dots l\}$. In general, as a consequence of the previous definitions, we will have,
\eq{\omega|_\Sigma\leq \d^l\sigma\sqrt{g|_\Sigma}~.}
The submanifold is called {\it calibrated with respect to} $\omega$ if it saturates the above inequality. 
Calibrated submanifolds then have the property that they minimize the volume 
within their homology class,
\eq{V_\Sigma=\int_\Sigma d^l\sigma\sqrt{g|_\Sigma}=\int_\Sigma\omega=\int_{\Sigma^\prime}\omega\leq \int_{\Sigma^\prime} d^l\sigma\sqrt{g|_{\Sigma^\prime}}
=V_{\Sigma^\prime}~,
}
where $\Sigma^\prime-\Sigma=\partial\widetilde{\Sigma}$ for some $\widetilde{\Sigma}$, and we have used Stokes theorem.

It has been known for some time \cite{calib1} that in the absence of flux supersymmetric 
branes (i.e. branes that do not break the supersymmetry of the background) can be thought of as extended objects wrapping calibrated submanifolds in spacetime. 
Therefore in the absence of flux supersymmetric branes are branes that minimize the volume in their homology class. In the presence of flux supersymmetric branes can be seen to minimize the energy in their generalized homology class.  
 
Let us review how this goes \cite{calib2,calib3}: We will consider in particular the following 
type II supergravity backgrounds. The ten-dimensional spacetime metric is of warped-product form,
\eq{
d s^2 = e^{2A}d s^2(\mathbb{R}^{1,d-1})+d s^2(\mathcal{M}_{10-d})
~,}
where $d=2,4,6,8$, and $\mathcal{M}_{10-d}$ is a Riemannian spin manifold. The RR flux is parameterized as follows,
\eq{F^{\mathrm{tot}}=\mathrm{vol}_d\wedge F^{\mathrm{el}}+F~,}
where $F$ is an even/odd polyform in IIA/IIB that lives in the internal space $\mathcal{M}_{10-d}$, and 
\eq{F^{\mathrm{el}}=\left(e^{A}\right)^d\star_{10-d}\sigma(F)~,}
implements the generalized self-duality condition in the democratic formalism for this particular type of backround. 
Consider a D-brane with worldvolume flux $\mathcal{F}$, extended in $q$ noncompact (external) spacetime directions, wrapping a submanifold $\Sigma$ in the internal space. At each point $p\in\Sigma$ its energy density is given by 
\eq{\label{ed}\mathcal{E}(T,\mathcal{F})
~=~
e^{qA-\phi}\sqrt{det(g|_{T}+\mathcal{F})}-\delta_{q,d}
\left(C^{\mathrm{el}}\wedge e^\mathcal{F}\right)_T~,}
where  $\phi$ is the dilaton,  $d\mathcal{F}=H|_\Sigma$, with $H$ the NS three-form, ${d}_HC^{\mathrm{el}}=F^{\mathrm{el}}$, and $T$ is the tangent space of $\Sigma$ at $p$. 

The generalized calibration form is defined to obey the following 
 conditions, in analogy to the ordinary case,
\eq{\spl{\label{gc}\mathrm{d}_H{\omega}~&=~\delta_{q,d}F^{\mathrm{el}}\\
\left({\omega}\wedge e^{\mathcal{F}}\right)_T
~&\leq~
 e^{qA-\phi}\sqrt{det(g+\mathcal{F})_T}
~,}}
for any oriented subspace $T$ with flux $\mathcal{F}$. The upshot is that 
generalized calibrated submanifolds, which are now defined as submanifolds which saturate the above inequality,  minimize the energy in their 
generalized homology class. Recall that two genreralized cycles $(\Sigma,\mathcal{F})$ are said to be in the same homology class if: (a)  $\Sigma^\prime-\Sigma=\partial\widetilde{\Sigma}$ for some $\widetilde{\Sigma}$, and (b)
 $\widetilde{\mathcal{F}}$ on $\widetilde{\Sigma}$ such that 
$\widetilde{\mathcal{F}}|_\Sigma=\mathcal{F}$ and  $\widetilde{\mathcal{F}}|_{\Sigma'}=\mathcal{F}'$. Then using Stokes theorem and the fact that $\d\mathcal{F}=H_{\Sigma}$ one obtains,
\eq{\spl{
\int_{\Sigma'} \mathrm{d}\sigma~\! \mathcal{E}(\Sigma',\mathcal{F}')&\geq
\int\left(\omega-\delta_{q,d}C^{\mathrm{el}}\right)_{\Sigma'}\wedge e^{\mathcal{F}'}\\
&=\int \left(\omega-\delta_{q,d}C^{\mathrm{el}}\right)_\Sigma\wedge e^\mathcal{F}
=\int_\Sigma \mathrm{d}\sigma ~\! \mathcal{E}(\Sigma,\mathcal{F})
~.}}
Moreover it can be shown that supersymmetric D-branes are calibrated as a
consequence of kappa symmetry. Indeed the kappa-symmetry projector $\g$ obeys 
an algebraic inequality of the form $|\epsilon|^2\geq\widetilde{\epsilon}\g\epsilon$, where $\epsilon$ is the background supersymmetry parameter. 
This can be rewritten schematically as follows:
\eq{\spl{\label{in}
&\sqrt{det(g+\mathcal{F})}
\geq\\
&~~~~~~~~~\frac{1}{|\epsilon|^2}
\sum_{2n+l=p+1}\frac{1}{2^nn!l!}\varepsilon^{a_1\dots a_{2n}b_1\dots b_l}
\mathcal{F}_{a_1a_2}\dots\mathcal{F}_{a_{2n-1}a_{2n}}\underbrace{\widetilde{\epsilon}\gamma_{b_1\dots b_l}\epsilon}_{|\epsilon|^2\omega}
~.}}
Identifying, via the Clifford map, the polyform $\omega$ as the spinor bilinear indicated in the equation above gives the second line of (\ref{gc}). 
The first equation in (\ref{gc}) is also satisfied  
as a consequence of the supersymmetry of the  background, i.e.~it follows from 
the Killing-spinor equation for $\epsilon$. Moreover taking into account that supersymmetric branes obey,
\eq{
(1-\gamma)\epsilon=0
~,}
thus saturating the inequality in (\ref{in}) and therefore also (\ref{gc}), it follows that supersymmetric branes 
are calibrated.

\begin{figure}
\centering
\includegraphics{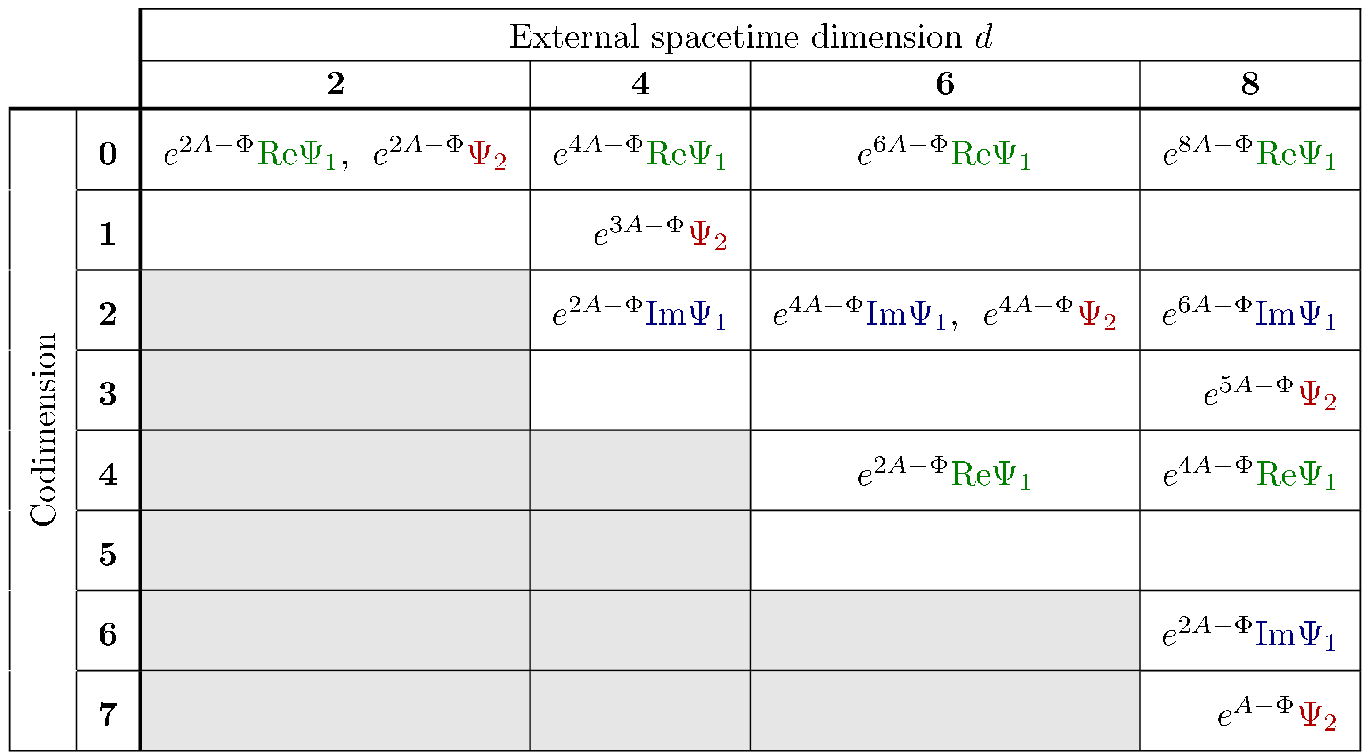}
\caption{\label{f6}The periodic table of generalized calibrations of type II supergravity for different dimensions of the external space $\mathbb{R}^{1,d-1}$. Reprinted from \cite{patalong}.}
\end{figure}

One then has a clear prescription of how to construct generalized calibrations, 
in correspondence with admissible supersymmetric branes in the background. This 
procedure can be carried out systematically for any external spacetime dimension $d$ and leads to the table of figure \ref{f6} for supersymmetric static, magnetic branes \cite{patalong}.\footnote{Static means that the D-branes extend in the time direction; magnetic means that their worldvolume flux is entirely along the spatial directions.}
The corresponding differential calibration conditions read,
\begin{align}\label{form}
\mathrm{d}_H\left( e^{(d-4r)A-\phi}{\color{mygreen}\mathrm{Re}\Psi_1}\right)&\color{myblack}=\delta_{r,0}F^{\mathrm{el}}~;
~~~~\!~~ \color{myblack}{d-4r\geq 1}\nn\\\color{myblack}
\mathrm{d}_H\left( e^{(d-2-4r)A-\phi}{\color{myblue}\mathrm{Im}\Psi_1}\right)&\color{myblack}=0~;
~~~~~~~~~\color{myblack}{d-2-4r\geq 1}\nn\\\color{myblack}
\mathrm{d}_H\left( e^{[\frac{1}{2}(d+2)-4r]A-\phi}{\color{myred}\Psi_2}\right)&\color{myblack}=0
~\!~;~~~\color{myblack}{\!\frac{1}{2}(d+2)-4r\geq 1}~;~~~r\in\mathbb{N}~.
\end{align}
Remarkably for certain spacetime dimensions it turns out that the content of these conditions is exactly equivalent to the supersymmetry conditions for the background. This was first pointed out in \cite{calib3} for the case $d=4$ \cite{gran} and subsequently  shown in \cite{patalong} for the case $d=6$. What happens in the case of other external spacetime dimensions? 
It is known that for backgrounds with generic supersymmetry parameters, not all background supersymmetry equations are of the form (\ref{form}) \cite{toma}. 
However if one restricts the supersymmetry parameters to be pure spinors, in other words: if one restricts to pure backgrounds (i.e.~backgrounds with $SU(n)\times SU(n)$ structure), it is not logically excluded that the differential conditions (\ref{form}) continue to be equivalent to the background supersymmetry conditions. 

As we will see in the following, for $d=2$ the supersymmetry/calibrations correspondence breaks down, although these backgrounds are still nicely described by GCG. More precisely, there is one `missing' pure spinor equation which is very similar to the ones above but is given in terms of the generalized $\d^c$ operator instead of the ordinary twisted differential. The situation will turn out to be even more complicated for $d=0$ (Euclidean ten-dimensional backgrounds of $SU(5)\times SU(5)$ structure), in which case not all background supersymmetry equations are of the form (\ref{form}).

Let us examine the $d=2$ case more closely. The spinor ansatz takes the form,
\eq{\label{introspindecompa2}
\epsilon_a=\zeta\otimes\eta_a+\zeta^c\otimes\eta_a^c
~, ~~~a=1,2~,}
where $\zeta$ is a complexified positive-chirality Killing spinor of $\mathbb{R}^{1,1}$
and the 
$\eta_a$'s are pure eight-dimensional spinors such that:
\eq{\label{normeta1}
|\eta_1|^2=|\eta_2|^2=\mathrm{const}\times e^{\frac{1}{2}A}~,
}
where $A$ is the warp factor. This equation can be seen to be equivalent to the 
requirement that the background admits calibrated branes. In this case the 
background susy equations which are in correspondence with 
D-brane calibrations take the form,
\boxedeq{\spl{\label{conjp}
  \d_H \left( e^{2 A- \Phi } \text{Re} \Psi_1 \right) &= 
F^{\mathrm{el}}\\
   \d_H \left( e^{2 A- \Phi} \Psi_2 \right) &= 0
~.}}
In particular the second equation above imples that the GACS $\mathcal{I}_2$,  
associated with the generalized pure spinor $\Psi_2$, is integrable. 
Moreover there is an extra differential equation which reads,
\boxedeq{\label{missing}
  \d^{\mathcal{I}_2}_H \left( e^{- \Phi } \text{Im} \Psi_1 \right) = F
~.}
This was shown for $SU(4)$-structure backgrounds in \cite{pt}, where it was conjectured that is should hold more generally and suggested a way to prove it in the general case. The general proof was given by Rosa in \cite{rosa}. 

We see that eqn.~(\ref{missing}) is expressed in terms of the generalized $\d^c$ operator  associated with the integrable GACS $\mathcal{I}_2$. This operator, that has been studied by Cavalcanti in \cite{cavalcanti}, appeared for the first time in supergravity in the $d=4$ case \cite{tomadol}, where it was used to reexpress one of the  background supersymmetry equations. 

For later use let us also give an equivalent form of the eqn.~(\ref{missing}) which is 
more readily generalizable to the $d=0$ case:
\eq{{\label{main1}
i\bar{\partial}^{\mathcal{I}_2}_H\big(e^{-\phi}\mathrm{Im}\Psi_1\big)=F^-
~,}}
where $F^-$ is the projection of $F$ onto $\sum_k\sum_{l \leq 0}\oplus ~\!U_{k,l}$.

Two particular cases of the system of equations (\ref{conjp}),(\ref{missing}) turn out to be related to those of \cite{Lau:2014fia}, as was recently shown in \cite{Minasian:2016txd}. More specifically, consider the `IIB complex system' given by the strict $SU(4)$ ansatz, $\eta_2 = e^{i \theta} \eta_1$, so that,
\eq{\spl{
\Psi_1 &= e^{-i \theta} e^{- i J} \\
\Psi_2 &= e^{i \theta} \Omega \;
~.}}
The solutions of this system were already studied in \cite{pt} and give rise to 
complex internal manifolds $\mcal_8$. The IIB complex system turns out to be a more general case of what was refered to in \cite{Minasian:2016txd} as the `2B LTY system' of \cite{Lau:2014fia}. 

Consider now the `IIB symplectic system' which is given by a different strict $SU(4)$ ansatz, $\eta_2 = e^{-i \theta} \eta^c_1$, so that the roles of $\Psi_{1,2}$ 
are interchanged,
\eq{\spl{
\Psi_1 &= e^{i \theta} \Omega \\
\Psi_2 &= e^{-i \theta} e^{- i J} 
~.}}
The solutions of this system give rise to 
symplectic internal manifolds $\mcal_8$ \cite{Minasian:2016txd}. The relation of the IIB symplectic system with what was refered to in \cite{Minasian:2016txd} as the `2A LTY system' of \cite{Lau:2014fia}, turns out to be a lot more complicated than 
 the relation between the IIB complex and the 2B LTY system.

The reason for this discrepancy was explained in \cite{Minasian:2016txd}: 
The 2A LTY, 2B LTY systems were educated guesses for what the susy conditions might look like in $d=2$ (eight internal dimensions), based on their form in $d=4$ (six internal dimensions). By construction they are mirror-symmetric with respect to the Fourier-Mukai transform, 
\eq{\label{t}
\!\!\!\!\!\!\!\mathrm{2A~LTY}\xleftrightarrow{~\mathrm{FM}{~~\!\!}}\mathrm{2B~LTY}
~.}
On the other hand the IIB complex and symplectic systems are T-dual in the 
generalized-geometric sense \cite{Cavalcanti:2011wu},
\eq{\label{u}
\mathrm{IIB~complex}\xleftrightarrow{~\mathrm{TD}{~~\!\!}}\mathrm{IIB~symplectic}~,
}
which corresponds to the interchange of $\Psi_1$ with $\Psi_2$.  
Recall that, roughly-speaking, the Fourier-Mukai transform is an isomorphism of the differential complexes, 
\eq{\label{q}
(\partial,\bar{\partial})\leftrightarrow(\d^{\Lambda},\d)
~,}
where $\d^\Lambda:=[\d,\Lambda]$ and $\Lambda$ is the adjoint of the Lefschetz operator $J\wedge$. 
On the other hand T-duality is an isomorphism,
\eq{\label{s}
(\d^{c},\d)\leftrightarrow(\d^{\Lambda},\d)
~.}
The difference between (\ref{q}),(\ref{s}) explains the 
aforementioned discrepancy.

Let us finally come to the case of zero external directions. Here we will consider  pure backgrounds of {\it Euclidean} ten-dimensional type II supergravity. More specifically, the spinor ansatz reads,
\eq{\label{introspindecompa1}
\epsilon_a=\eta_a~, ~~~a=1,2~,}
where the $\eta_a$'s are pure and,
\eq{\label{normeta2}
|\eta_1|^2=|\eta_2|^2=\alpha^2~
,}
which is imposed by analogy to the $d\geq 2$ case, with 
$\alpha$ a function on $\mcal_{10}$. As was shown in \cite{Prins:2014ona}, in this 
case most of the susy equations can be packaged in a system of equations which is directly analogous to the $d\geq2$ cases,
\boxedeq{\spl{\label{main}
\d_H\big(\alpha^2e^{-\phi} \Psi_2\big)&=0\\
i\bar{\partial}^{\mathcal{I}_2}_H\big(e^{-\phi}\mathrm{Im}\Psi_1\big)&=F^-
~.}}
The first of these equations was proven in general, whereas the second was only 
shown in \cite{Prins:2014ona} to hold for strict $SU(5)$ type II backgrounds. The fact that this equation is manifestly mirror-symmetric (in the sense that when written in terms of generalized pure spinors it takes the same form in IIA as it does in IIB) is a very strong indication that it should hold for general $SU(5)\times SU(5)$ backgrounds. 

Moreover there is a (small) number of `missing' equations which do not seem natural from the GCG point of view. For the strict $SU(5)$ case these read (see \cite{Prins:2014ona} for further details on the notation):
\eq{f_4^{(1,0)} = - \frac{1}{2} e^{-\phi} e^{i\theta} 
\left( \partial^+ \log\alpha + i h^{(1,0)}\right)
~,}
in IIA, and 
\eq{\spl{
e^\phi e^{i\theta} \big( \frac{1}{4} f_{1}^{(1,0)} + i f_{3}^{(1,0)} - \frac{3}{2} f_{5}^{(1,0)} \big)
&= -{\partial^+}\big( 2 \log \alpha  -\frac{i}{2}\theta \big)- i h^{(1,0)} \\
e^\phi e^{- i\theta} \big(- \frac{1}{4} f_{1}^{(1,0)} + i f_{3}^{(1,0)} +\frac{3}{2} f_{5}^{(1,0)} \big)
&= -{\partial^+}\big( 2 \log \alpha  +\frac{i}{2}\theta \big)+ i h^{(1,0)} 
~,}}
in IIB. Note however that these extra equations are simply constraints on the 
fluxes, i.e. they do not impose any additional restrictions on the geometry of  $\mcal_{10}$.

\subsection{Conclusions}

Supersymmetric pure backgrounds of type II supergravity 
(equivalently $SU(n)\times SU(n)$-structure backgrounds) of the form 
$\mathbb{R}^{1,d-1}\times\mcal_{2n}$ are characterized by 
pairs of pure Killing spinors of the internal manifold $\mcal_{2n}$. 
The study of these backgrounds  as a function of the external spacetime dimension $d$ reveals several general patterns and 
periodicities which are not visible for fixed $d$.

These backgrounds have a natural and complete description within GCG for $d\geq 2$. For the case of Euclidean type II supergravity, $d=0$, the geometric constraints for supersymmetric pure backgrounds are also naturaly described in GCG. However in that case there are 
a few additional constraints imposed by supersymmetry on the fluxes, but not on the geometry of $\mcal_{10}$, which do not seem natural from the point of view of GCG. 
 Moreover the proof of the second equation in (\ref{main})  has not yet been given in  the general $SU(5)\times SU(5)$ case. Nevertheless, because 
of its mirror-symmetric form, there is very good reason to believe it is valid beyond the strict $SU(5)$ case.

One interesting general feature that emerges from these results is that the generalized Dolbeault operator plays a fundamental role in generic external spacetime dimensions, a fact which was not fully appreciated in the early works on the subject.  It has been suggested in \cite{tomadol}  that the formulation in terms of the generalized Dolbeault operator should facilitate the search for existence theorems for  
supersymmetric flux vacua.

The supersymmetry/calibrated D-branes correspondence which holds for $d\geq 4$ breaks down in lower external spacetime dimensions. It would be nice to have a better understanding of this phenomenon: is there another principle that replaces the correspondence?

\acknowledgments

I am grateful to the organisers of the Corfu Summer School 2014 `Quantum Fields and Strings'  
and the  May 2016 `Generalized Geometry \& T-dualities' conference at the Simons Center for Geometry and Physics for hospitality and financial support.

\begin{appendix}

\section{Spinors and gamma matrices in Euclidean spaces}\label{sec:spinors}

In this section we list some useful relations and 
explain in more detail our spinor conventions for general even-dimensional Euclidean spaces of dimension $D=2n$. 

The charge conjugation matrix obeys:
\eq{\label{c}
C^{\mathrm{Tr}}=(-1)^{\frac{1}{2}n(n+1)}C~;~~C^*=(-1)^{\frac{1}{2}n(n+1)}C^{-1}~;~~
\gamma_m^{\mathrm{Tr}}=(-1)^n C^{-1}\gamma_m C
~.
}
The chirality matrix $\gamma_{2n+1}$ is defined by:
\eq{\label{chirdef}
\gamma_{2n+1}:=i^n\gamma_1\dots \gamma_{2n}
~,}
and obeys
\eq{\label{chirob}
\gamma^{\mathrm{Tr}}_{2n+1}
=(-1)^n C^{-1}\gamma_{2n+1}C
~,}
as follows from eqs.~(\ref{chirdef}, \ref{c}). The chirality projector: 
\eq{
P_{\pm}:=\frac12(1\pm\gamma_{2n+1})
~,} 
projects a Dirac spinor $\chi$ onto the chiral, antichiral Weyl parts $\chi_\pm$:
\eq{
\chi_\pm=P_\pm\chi~.
}
Taking eq.~(\ref{chirob}) into account we obtain:
\eq{\label{ptr}
C^{-1}P_\pm=
\left\{
\begin{array}{ll}
P_\pm^{\mathrm{Tr}}C^{-1}~,& ~~~n=\mathrm{even}\\
P_\mp^{\mathrm{Tr}}C^{-1}~,& ~~~n=\mathrm{odd}\\
\end{array}
\right.
~.}
Covariantly-transforming spinor bilinears must be of the form $(\widetilde{\psi}\gamma_{m_1\dots m_p}\chi)$, 
where in any dimension we define:
\eq{
\widetilde{\psi}:=\psi^{\mathrm{Tr}}C^{-1}
~.}
The complex conjugate $\eta^c$ of a spinor $\eta$ is given by:
\eq{
\eta^c:=C\eta^*
~,}
form which it follows that:
\eq{\label{a3}
(\eta^c)^c=(-1)^{\frac12 n(n+1)}\eta~;~~~\eta^{\dagger}=
(-1)^{\frac{1}{2}n(n+1)}\widetilde{\eta^c}
~.}
Using eq.~(\ref{ptr}) we find:
\eq{\spl{\label{bilins}
(\widetilde{\psi}_\pm\gamma_{m_1\dots m_{2l}}\chi_\mp)&=0=(\widetilde{\psi}_\pm\gamma_{m_1\dots m_{2l+1}}\chi_\pm)~,~~n=\mathrm{even}\\
(\widetilde{\psi}_\pm\gamma_{m_1\dots m_{2l}}\chi_\pm)&=0=(\widetilde{\psi}_\pm\gamma_{m_1\dots m_{2l+1}}\chi_\mp)~,~~n=\mathrm{odd}
~.}}
Moreover:
\eq{\label{bilinstr}
(\widetilde{\psi}\gamma_{m_1\dots m_p}\chi)=
(-1)^{np+\frac12 n(n+1)}(\widetilde{\chi}\gamma_{m_p\dots m_1}\psi)
=
(-1)^{\frac12 (n-p)(n-p+1)}(\widetilde{\chi}\gamma_{m_1\dots m_p}\psi)
~,}
which can be shown using e.g. the identity,
\eq{\label{a1}
\gamma_{m_1\dots m_p}^{\mathrm{Tr}}=(-1)^{np}(-1)^{\frac12 p(p-1)}C^{-1}\gamma_{m_1\dots m_p}C
~.}
Moreover, the identity
\eq{\label{a2}
\gamma_{m_1\dots m_p}^*=(-1)^{np}C^{-1}\gamma_{m_1\dots m_p}C
~,}
can be used to show the following relations:
\eq{\spl{\label{bilinscomp}
(\widetilde{\psi}\gamma_{m_1\dots m_p}\chi)^*&
=(-1)^{np}(\widetilde{\psi^c}\gamma_{m_1\dots m_p}\chi^c)\\
(\widetilde{\psi}\gamma_{m_1\dots m_p}\chi^c)^*&
=(-1)^{np+\frac12 n(n+1)}(\widetilde{\psi^c}\gamma_{m_1\dots m_p}\chi)
~.}}

\end{appendix}

\end{document}